\documentclass[nofootinbib,preprint,prX]{revtex4-1} 

\usepackage{amsmath}
\usepackage{mathrsfs}
\usepackage{amssymb}		
\usepackage{graphicx}		
\usepackage{graphics}
\usepackage[small,compact]{titlesec}
\usepackage{axodraw}
\usepackage{xspace}
\usepackage{natbib}
\usepackage{colordvi}
\usepackage{units}

\newcommand{\vect}[1]{\vec{\mathbf{#1}}}
\newcommand{\vectS}[1]{\vec{\boldsymbol{#1}}}

\newcommand{\gThreeSOneThreePOne}{g^{{}^{3}\!S_{1}-{}^{3}\!P_{1}}}
\newcommand{\gThreeSOneOnePOne}{g^{{}^{3}\!S_{1}-{}^{1}\!P_{1}}}
\newcommand{\gDeltaIZero}{g^{{}^{1}\!S_{0}-{}^{3}\!P_{0}}_{(\Delta I=0)}}
\newcommand{\gDeltaIOne}{g^{{}^{1}\!S_{0}-{}^{3}\!P_{0}}_{(\Delta I=1)}}

\newcommand{\darrow}[1]{\stackrel{\leftrightarrow\!}{#1}}
\newcommand{\propagator}[2]{\frac{i}{#1-\frac{\vect{#2}^{2}}{2M_{N}}+i\epsilon}}

\newcommand{\EFT}{$\mathrm{EFT}_{\not{\pi}}$\xspace}

\begin{document}

\title{Parity Violation in \emph{nd} Interactions}

\author{Jared Vanasse}
\email{jjvanass@physics.umass.edu}
\affiliation{\emph{Department of Physics-LGRT,\\
University of Massachusetts,\\
Amherst, MA 01003}
}

\date{\today}

\begin{abstract}
 We calculate the parity-violating amplitudes in the $nd$ interaction with pionless effective field theory to LO.  Matching the parity violating low energy constants to the DDH coefficients we make numerical predictions for parity-violating observables.  In particular we give predictions for the spin rotation of a neutron on a deuteron target, and target and beam asymmetries in $nd$ scattering.
\end{abstract}


\maketitle

\section{Introduction}
Hadronic parity-violation has been traditionally analyzed in terms of potential models; specifically the DDH model\cite{Desplanques:1979hn}, which is a parity
violating single meson exchange picture containing seven phenomenological constants.
 However, there exist well known discrepancies between experimental measurements and the DDH
 model\cite{Holstein:2010zza}.  Some of this discrepancy is no doubt due to
 nuclear physics uncertainties, but another source may be the use of the
 model-dependent DDH potential.  A possible solution to these problems has recently been
 proposed by Zhu et al.\cite{Holstein:2010zza,Zhu:2004vw}, restriction of experiments to nuclei with
 $A<4$ so that nuclear uncertainties are minimal, and analysis using a
 model-independent picture via effective field theory.  At low energies, less than $m_{\pi}^{2}/M_{N}\sim 20 MeV$, such
 an approach is provided by Pionless EFT (\EFT), which has been extremely successful
 at low energies in the two-body and three-body sector for parity-conserving
 interactions, including interactions with external currents\cite{Beane:2000fx}.  At low energies inclusion of parity-violation requires only five additional low energy constants (LEC's) in the nucleon-nucleon
 interaction.  These LEC's involve all possible isospin structures that mix S
 and P waves with one derivative and are equivalent to the parameters originally posited by Danilov\cite{Danilov}.  The fact that five LEC's are needed has also been specifically shown by
 Girlanda, who does this by performing a non-relativistic reduction of all
 possible one derivative relativistic parity-violating structures that conserve
 CP\cite{Girlanda:2008ts}.  Calculations for parity-violation using EFT methods
 have heretofore been primarily focused in the two-body sector.  Such
 calculations include parity violation in nucleon-nucleon scattering and in the
 process $np \to d\gamma$\cite{Phillips:2008hn,Schindler:2009wd,Shin:2009hi}. 
 Parity-violating EFT calculations have been done in the three-body sector using
 a hybrid approach, wherein the parity-violating potential is given by \EFT, but
 is used with wavefunctions determined by either the variational method or by a
 Fadeev integral equation technique\cite{Schiavilla:2008ic,PhysRevC.83.029902,Song:2010sz}.  Such
 calculations include neutron spin rotation and beam asymmetry in $nd$ interactions.  Recently a paper by Griesshammer, Schindler, and Springer predicted the spin rotation of a neutron on a deuterium target up to and including NLO effects in \EFT\!\!\cite{Griesshammer:2011md}.  However, they only included order of magnitude estimates for the parity violating coefficients and left open the calculation of other possible parity violating observables in $nd$ interactions.  In this paper we set out to obtain estimates for the parity violating coefficients by matching them to the DDH ``best" value estimates.  As well as calculating  the neutron spin rotation on a deuteron target at LO in \EFT, we also calculate the beam and target asymmetry at LO in $nd$ scattering. Estimation of these observables will then allow one to assess the feasibility of $nd$ interactions as a realistic experimental probe for the five PV LEC's.  Below we calculate the LO amplitudes for S-P mixing in $nd$ scattering due to the two-body parity-violating Lagrangian.  (Since Griesshammer and Schindler showed that no three-body parity-violating force occurs up to and including NLO, only five LEC's exist at LO\cite{Griesshammer:2010nd}.)  Predictions are made for PV observables and numerical estimates are given based on DDH ``best" value estimates.  In a future publication we shall present higher order corrections.  The paper is organized as follows.  In section II we give the form of the two-body parity-violating interaction.  Then in section III we show what diagrams are needed at LO and how to calculate them.  Section IV shows how estimates for parity-violating LEC's can be obtained, and in Section V we show how to relate our amplitudes to observables.  Finally in Section VI we summarize the results.
\vspace{2cm}

\section{Two-Body Parity-Violating Interaction}

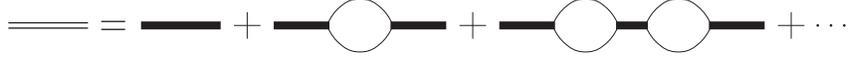
\begin{figure}
\begin{center}\begin{picture}(325,56)(90,-10)
\Line(100,-1)(130,-1)
\Line(100,1)(130,1)
\Line(135,1.5)(144,1.5)
\Line(135,-1.5)(144,-1.5)
\SetWidth{3}
\Line(150,0)(180,0)
\SetWidth{.3}
\Line(185,0)(195,0)
\Line(190,-5)(190,5)
\SetWidth{3}
\Line(200,0)(221,0)
\SetWidth{.3}
\Curve{(220,0)(221,2)(226.25,8)(232.5,10)(238.75,8)(244,2)(245,0)}
\Curve{(220,0)(221,-2)(226.25,-8)(232.5,-10)(238.75,-8)(244,-2)(245,0)}
\SetWidth{3}
\Line(244,0)(265,0)
\SetWidth{.3}
\Line(270,0)(280,0)
\Line(275,-5)(275,5)
\SetWidth{3}
\Line(285,0)(306,0)
\SetWidth{.3}
\Curve{(305,0)(306,2)(311.25,8)(317.5,10)(323.75,8)(329,2)(330,0)}
\Curve{(305,0)(306,-2)(311.25,-8)(317.5,-10)(323.75,-8)(329,-2)(330,0)}
\SetWidth{3}
\Line(329,0)(341,0)
\SetWidth{.3}
\Curve{(340,0)(341,2)(346.25,8)(352.5,10)(358.75,8)(364,2)(365,0)}
\Curve{(340,0)(341,-2)(346.25,-8)(352.5,-10)(358.75,-8)(364,-2)(365,0)}
\SetWidth{3}
\Line(364,0)(385,0)
\SetWidth{.3}
\Line(390,0)(400,0)
\Line(395,-5)(395,5)
\SetWidth{3}
\Vertex(405,0){.5}
\Vertex(410,0){.5}
\Vertex(415,0){.5}
\end{picture}\end{center}
\caption{\label{fig:LODeuteronPropagator} Infinite sum of nucleon bubbles contributes to LO deuteron propagator}
\end{figure}

The leading order two-body parity-conserving Lagrangian in the auxiliary field formalism is given by

\begin{align}
\mathcal{L}_{PC}^{d}=\ &N^{\dagger}\left(i\partial_{0}+\frac{\vect{\nabla}^2}{2M_{N}}\right)N-t_{i}^{\dagger}\left(i\partial_{0}+\frac{\vect{\nabla}^{2}}{4M_{N}}-\Delta^{({}^3\!S_{1})}_{(-1)}-\Delta^{({}^3\!S_{1})}_{(0)}\right)t_{i}+y_{d}\left[t_{i}^{\dagger}N ^{T}P_{i}N +h.c.\right]\\\nonumber
&-s_{a}^{\dagger}\left(i\partial_{0}+\frac{\vect{\nabla}^{2}}{4M_{N}}-\Delta^{({}^1\!S_{0})}_{(-1)}-\Delta^{({}^1\!S_{0})}_{(0)}\right)s_{a}+y_{t}\left[s_{a}^{\dagger}N^{T}\bar{P}_{a}N+h.c.\right]
\end{align}

\noindent where $t_{i}$ ($s_{a}$) is the deuteron (singlet) auxiliary field\cite{Gabbiani:1999yv}.  Here $P_{i}=\frac{1}{\sqrt{8}}\sigma_{2}\sigma_{i}\tau_{2}$ projects out the ${}^{3}\!{S}_{1}$ channel and $\bar{P}_{a}=\frac{1}{\sqrt{8}}\sigma_{2}\tau_{2}\tau_{a}$ projects out the ${}^{1}\!S_{0}$ channel.   The auxiliary field formalism is equivalent to the partial wave formalism in which only nucleon fields are used as can be seen by integrating over the auxiliary fields and using a field redefinition\cite{Bedaque:1999vb}. In this formulation the two-body parity-violating Lagrangian amplitude is given by a form including five low energy constants\cite{Schindler:2009wd}.

\begin{align}
\label{eq:AuxLagr}
\mathcal{L}^{d}_{PV}=-&\left[g^{({}^{3}\!S_{1}-{}^{1}\!P_{1})}t_{i}^{\dagger}\left(N^{t}\sigma_{2}\tau_{2}i\darrow{\nabla}_{i}N\right)\right.\\\nonumber
&+g^{({}^{1}\!S_{0}-{}^{3}\!P_{0})}_{(\Delta I=0)}s_{a}^{\dagger}\left(N^{T}\sigma_{2}\vectS{\sigma}\cdot\tau_{2}\tau_{a}i\darrow{\nabla}N\right)\\\nonumber
&+g^{({}^{1}\!S_{0}-{}^{3}\!P_{0})}_{(\Delta I=1)}\epsilon^{3ab}(s^{a})^{\dagger}\left(N^{T}\sigma_{2}\vectS{\sigma}\cdot\tau_{2}\tau^{b}\darrow{\nabla}N\right)\\\nonumber
&+g^{({}^{1}\!S_{0}-{}^{3}\!P_{0})}_{(\Delta I=2)}\mathcal{I}^{ab}(s^{a})^{\dagger}\left(N^{T}\sigma_{2}\vectS{\sigma}\cdot\tau_{2}\tau^{b}i\darrow{\nabla}N\right)\\\nonumber
&+\left.g^{({}^{3}\!S_{1}-{}^{3}\!P_{1})}\epsilon^{ijk}(t_{i})^{\dagger}\left(N^{T}\sigma_{2}\sigma^{k}\tau_{2}\tau_{3}\darrow{\nabla}^{j}N\right)\right]+h.c.
\end{align}


\noindent where $a\darrow{\nabla}b=a(\overrightarrow{\nabla})b-(\overrightarrow{\nabla}a)b$, and ${\mathcal{I}}=\mathrm{diag}[1,1,-2]$ projects out the isotensor contribution.\newline\indent  The deuteron kinetic energy and the term $\Delta_{(0)}^{({}^3\!S_{1})}$ are sub-leading with respect to $\Delta_{(-1)}^{({}^3\!S_{1})}$.  Thus at LO the bare deuteron propagator is given by $i/\Delta_{(-1)}^{({}^3\!S_{1})}$ which is then dressed by an infinite number of nucleon bubbles as seen inf Fig. \ref{fig:LODeuteronPropagator}.  The resulting propagator depends on $\Delta_{(-1)}^{{}^{3}\!S_{1}}$ and $y_{d}^{2}$, with values adjusted such that the deuteron propagator has its pole at the correct value.  A similar calculation can be carried out for the propagator of the singlet auxiliary field.  This procedure has been carried out in many papers, the end results for the LO deuteron and singlet propagator are listed below\cite{Gabbiani:1999yv}.  Also we include the constraints imposed on the coefficients $\Delta_{(-1)}^{{}^{3}\!S_{1}}$, $y_{d}^{2}$, and their ${}^{1}\!S_{0}$ counterparts at LO.  Note the presence of the parameter $\mu$, which is a cutoff imposed by using dimensional regularization with the PDS subtraction scheme\cite{Kaplan:1998tg}. (Here $\gamma_{d}=45.7025$ MeV is the deuteron binding momentum and $\gamma_{t}=1/a_{t}$, where $a_{t}=-23.714$ fm is the scattering length in the ${}^{1}\!S_{0}$ channel.)

\begin{equation}
iD_{d}(p_{0},\vect{p})=\frac{4\pi i}{M_{N}y_{d}^{2}}\frac{1}{\gamma_{d}-\sqrt{\frac{\vect{p}^{2}}{4}-M_{N}p_{0}-i\epsilon}}
\end{equation}

\begin{equation}
iD_{t}(p_{0},\vect{p})=\frac{4\pi i}{M_{N}y_{t}^{2}}\frac{1}{\gamma_{t}-\sqrt{\frac{\vect{p}^{2}}{4}-M_{N}p_{0}-i\epsilon}}
\end{equation}

\begin{equation}
\frac{\Delta_{(-1)}^{{}^{3}\!S_{1}}}{y_{d}^{2}}=\frac{M_{N}}{4\pi}(\gamma_{d}-\mu),\quad\frac{\Delta_{(-1)}^{{}^{1}\!S_{0}}}{y_{t}^{2}}=\frac{M_{N}}{4\pi}(\gamma_{t}-\mu)
\end{equation}
\vspace{.2cm}

\section{Three-Body Parity Violation LO}

\begin{figure}[hbt]

\begin{center}\begin{picture}(470,60)(90,-20)
\SetScale{1.15}
\Line(100,25.5)(160,25.5)
\Line(100,23.5)(160,23.5)
\ArrowLine(100,-5)(130,-5)
\ArrowLine(130,-5)(160,-5)
\COval(130,10)(16.5,10)(0){Black}{Yellow}
\PText(130,10)(0)[c]{PV}

\Line(170,10)(179,10)
\Line(170,12.5)(179,12.5)

\Line(190,25.5)(210,25.5)
\Line(190,23.5)(210,23.5)
\ArrowLine(210,24.5)(240,24.5)
\ArrowLine(190,-5)(220,-5)
\Line(220,-6)(240,-6)
\Line(220,-4)(240,-4)
\ArrowLine(210,24.5)(220,-5)
\CBoxc(210,24.5)(5,5){Green}{Green}

\Line(250,10)(260,10)
\Line(255,5)(255,15)

\Line(270,25.5)(300,25.5)
\Line(270,23.5)(300,23.5)
\SetColor{Blue}
\SetWidth{3}
\Line(300,24.5)(320,24.5)
\SetWidth{.5}
\SetColor{Black}
\ArrowLine(320,24.5)(350,24.5)
\ArrowLine(270,-5)(300,-5)
\ArrowLine(300,-5)(330,-5)
\Line(330,-4)(350,-4)
\Line(330,-6)(350,-6)
\ArrowLine(320,24.5)(330,-4.5)
\COval(300,10)(16.5,10)(0){Black}{Red}
\CBoxc(320,24.5)(5,5){Green}{Green}
\PText(300,10)(0)[c]{PC}

\Line(360,10)(370,10)
\Line(365,5)(365,15)

\Line(380,25.5)(410,25.5)
\Line(380,23.5)(410,23.5)
\SetColor{Blue}
\SetWidth{3}
\Line(410,24.5)(430,24.5)
\SetWidth{.5}
\SetColor{Black}
\ArrowLine(430,24.5)(460,24.5)
\ArrowLine(380,-5)(410,-5)
\ArrowLine(410,-5)(440,-5)
\Line(440,-4)(460,-4)
\Line(440,-6)(460,-6)
\ArrowLine(430,24.5)(440,-4.5)
\COval(410,10)(16.5,10)(0){Black}{Yellow}
\PText(410,10)(0)[c]{PV}

\Line(250,-40)(260,-40)
\Line(255,-45)(255,-35)

\Line(270,-24.5)(300,-24.5)
\Line(270,-26.5)(300,-26.5)
\SetColor{Blue}
\SetWidth{3}
\Line(305,-25.5)(320,-40.5)
\SetWidth{.5}
\SetColor{Black}
\Line(320.5,-40)(335.5,-55)
\Line(318.5,-41)(332.5,-55)
\ArrowLine(305,-55)(320,-40)
\ArrowLine(320,-40)(335,-25)
\ArrowLine(270,-55)(300,-55)
\COval(300,-40)(16.5,10)(0){Black}{Yellow}
\PText(300,-40)(0)[c]{PV}

\end{picture}\end{center}

\begin{center}\begin{picture}(470,100)(90,-50)
\SetScale{1.15}
\Line(100,25.5)(130,25.5)
\Line(100,23.5)(130,23.5)
\DashLine(130,25.5)(160,25.5){3}
\DashLine(130,23.5)(160,23.5){3}
\ArrowLine(100,-5)(130,-5)
\ArrowLine(130,-5)(160,-5)
\COval(130,10)(16.5,10)(0){Black}{Yellow}
\PText(130,10)(0)[c]{PV}

\Line(170,10)(179,10)
\Line(170,12.5)(179,12.5)

\Line(190,25.5)(210,25.5)
\Line(190,23.5)(210,23.5)
\ArrowLine(210,24.5)(240,24.5)
\ArrowLine(190,-5)(220,-5)
\DashLine(220,-6)(240,-6){3}
\DashLine(220,-4)(240,-4){3}
\ArrowLine(210,24.5)(220,-5)
\CBoxc(210,24.5)(5,5){Green}{Green}

\Line(250,10)(260,10)
\Line(255,5)(255,15)

\Line(270,25.5)(320,25.5)
\Line(270,23.5)(320,23.5)
\SetColor{Blue}
\SetWidth{3}
\Line(300,24.5)(320,24.5)
\SetWidth{.5}
\SetColor{Black}
\ArrowLine(320,24.5)(350,24.5)
\ArrowLine(270,-5)(300,-5)
\ArrowLine(300,-5)(330,-5)
\DashLine(330,-4)(350,-4){3}
\DashLine(330,-6)(350,-6){3}
\ArrowLine(320,24.5)(330,-4.5)
\COval(300,10)(16.5,10)(0){Black}{Red}
\CBoxc(320,24.5)(5,5){Green}{Green}
\PText(300,10)(0)[c]{PC}

\Line(360,10)(370,10)
\Line(365,5)(365,15)

\Line(380,25.5)(410,25.5)
\Line(380,23.5)(410,23.5)
\SetColor{Blue}
\SetWidth{3}
\Line(410,24.5)(430,24.5)
\SetWidth{.5}
\SetColor{Black}
\ArrowLine(430,24.5)(460,24.5)
\ArrowLine(380,-5)(410,-5)
\ArrowLine(410,-5)(440,-5)
\DashLine(440,-4)(460,-4){3}
\DashLine(440,-6)(460,-6){3}
\ArrowLine(430,24.5)(440,-4.5)
\COval(410,10)(16.5,10)(0){Black}{Yellow}
\PText(410,10)(0)[c]{PV}

\Line(250,-40)(260,-40)
\Line(255,-45)(255,-35)

\Line(270,-24.5)(300,-24.5)
\Line(270,-26.5)(300,-26.5)
\SetColor{Blue}
\SetWidth{3}
\Line(305,-25.5)(320,-40.5)
\SetWidth{.5}
\SetColor{Black}
\DashLine(320.5,-40)(335.5,-55){3}
\DashLine(318.5,-41)(332.5,-55){3}
\ArrowLine(305,-55)(320,-40)
\ArrowLine(320,-40)(335,-25)
\ArrowLine(270,-55)(300,-55)
\COval(300,-40)(16.5,10)(0){Black}{Yellow}
\PText(300,-40)(0)[c]{PV}

\end{picture}\end{center}
\vspace{.2cm}
\caption{Integral Equations for Parity Violation at LO (Note diagrams where lower vertices are parity-violating are not included)}
\label{fig:LO_Integral_Equations}
\end{figure}
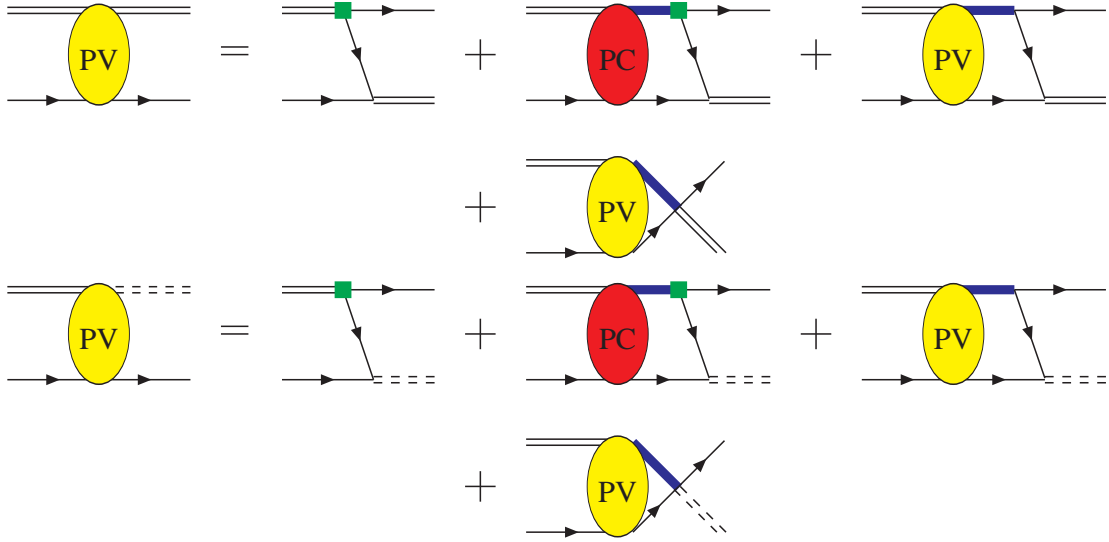

As in the parity-conserving case for three-body interactions one needs to solve an infinite sum of diagrams for the parity-violating amplitude at LO\cite{Gabbiani:1999yv,Bedaque:2002yg}, leading to a coupled set of integral equations.  Numerical solution is necessary, as such integral equations cannot be solved analytically.  In general we must solve a set of four coupled integral equations.  However, since parity-violation is so small, $G_{F}m_{\pi}^{2}\sim 10^{-7}$ we can ignore second order PV terms.  Then the integral equations for the parity-conserving amplitudes decouple\cite{Gudkov:2010pt}, and are exactly the same as in previous papers\cite{Gabbiani:1999yv,Griesshammer:2004pe,Bedaque:2002yg}. The remaining coupled parity-violating integral equations at LO are shown in Fig. \ref{fig:LO_Integral_Equations}, where the boxes represent parity-violating vertices, the double line the dressed deuteron propagator, the double dashed lines the dressed singlet propagator, and the line with arrow the nucleon propagator.  The thick lines represent a sum over both deuteron and singlet propagators.  Thus the thick line allows one to represent two Feynman diagrams with a single diagram.  There are also diagrams where two dibaryon lines and two nucleon lines meet at a single vertex, due to the three-body force term in the Lagrangian.

 This three-body force term enters at LO, only in the Doublet S-wave channel. (Note we have not yet projected out any specific channel.)  The momentum integrals are regulated 
 
\begin{figure}[hbt]
\begin{center}\begin{picture}(350,200)(160,-160)
\SetScale{1.2}
\Line(100,25.5)(130,25.5)
\Line(100,23.5)(130,23.5)
\Line(130,25.5)(160,25.5)
\Line(130,23.5)(160,23.5)
\ArrowLine(100,-5)(130,-5)
\ArrowLine(130,-5)(160,-5)
\COval(130,10)(16.5,10)(0){Black}{Yellow}
\PText(130,10)(0)[c]{PV}

\Line(170,10)(179,10)
\Line(170,12.5)(179,12.5)
\Text(175,10)[c]{$=$}

\Line(190,25.5)(210,25.5)
\Line(190,23.5)(210,23.5)
\ArrowLine(210,24.5)(240,24.5)
\ArrowLine(190,-5)(220,-5)
\Line(220,-6)(240,-6)
\Line(220,-4)(240,-4)
\ArrowLine(210,24.5)(220,-5)
\CBoxc(210,24.5)(5,5){Green}{Green}

\Line(250,10)(260,10)
\Line(255,5)(255,15)

\Line(270,25.5)(320,25.5)
\Line(270,23.5)(320,23.5)
\ArrowLine(320,24.5)(350,24.5)
\ArrowLine(270,-5)(300,-5)
\ArrowLine(300,-5)(330,-5)
\Line(330,-4)(350,-4)
\Line(330,-6)(350,-6)
\ArrowLine(320,24.5)(330,-4.5)
\COval(300,10)(16.5,10)(0){Black}{Red}
\CBoxc(320,24.5)(5,5){Green}{Green}
\PText(300,10)(0)[c]{PC}

\Line(360,10)(370,10)
\Line(365,5)(365,15)

\Line(380,25.5)(410,25.5)
\Line(380,23.5)(410,23.5)
\DashLine(410,25.5)(430,25.5){3}
\DashLine(410,23.5)(430,23.5){3}
\ArrowLine(430,24.5)(460,24.5)
\ArrowLine(380,-5)(410,-5)
\ArrowLine(410,-5)(440,-5)
\Line(440,-4)(460,-4)
\Line(440,-6)(460,-6)
\ArrowLine(430,24.5)(440,-4.5)
\COval(410,10)(16.5,10)(0){Black}{Red}
\CBoxc(430,24.5)(5,5){Green}{Green}
\PText(410,10)(0)[c]{PC}

\Line(250,-40)(260,-40)
\Line(255,-45)(255,-35)

\Line(270,-24.5)(290,-24.5)
\Line(270,-26.5)(290,-26.5)
\ArrowLine(270,-55)(300,-55)
\ArrowLine(290,-25.5)(300,-55)
\Line(290,-24.5)(350,-24.5)
\Line(290,-26.5)(350,-26.5)
\ArrowLine(300,-55)(320,-55)
\ArrowLine(320,-55)(350,-55)
\COval(320,-40)(16.5,10)(0){Black}{Red}
\CBoxc(290,-25.5)(5,5){Green}{Green}
\PText(320,-40)(0)[c]{PC}

\Line(360,-40)(370,-40)
\Line(365,-45)(365,-35)

\Line(380,-24.5)(400,-24.5)
\Line(380,-26.5)(400,-26.5)
\ArrowLine(380,-55)(410,-55)
\ArrowLine(400,-25.5)(410,-55)
\DashLine(400,-24.5)(430,-24.5){3}
\DashLine(400,-26.5)(430,-26.5){3}
\Line(430,-24.5)(460,-24.5)
\Line(430,-26.5)(460,-26.5)
\ArrowLine(410,-55)(430,-55)
\ArrowLine(430,-55)(460,-55)
\COval(430,-40)(16.5,10)(0){Black}{Red}
\CBoxc(400,-25.5)(5,5){Green}{Green}
\PText(430,-40)(0)[c]{PC}

\Line(190,-90)(200,-90)
\Line(195,-95)(195,-85)

\Line(210,-74.5)(260,-74.5)
\Line(210,-76.5)(260,-76.5)
\ArrowLine(260,-75.5)(290,-75.5)
\ArrowLine(210,-105)(240,-105)
\ArrowLine(240,-105)(270,-105)
\Line(270,-104)(320,-104)
\Line(270,-106)(320,-106)
\ArrowLine(260,-75.5)(270,-104.5)
\ArrowLine(290,-75.5)(320,-75.5)
\COval(240,-90)(16.5,10)(0){Black}{Red}
\COval(290,-90)(16.5,10)(0){Black}{Red}
\CBoxc(260,-75.5)(5,5){Green}{Green}
\PText(240,-90)(0)[c]{PC}
\PText(290,-90)(0)[c]{PC}

\Line(330,-90)(340,-90)
\Line(335,-95)(335,-85)

\Line(350,-74.5)(400,-74.5)
\Line(350,-76.5)(400,-76.5)
\ArrowLine(400,-75.5)(430,-75.5)
\ArrowLine(350,-105)(380,-105)
\ArrowLine(380,-105)(410,-105)
\DashLine(410,-104)(430,-104){3}
\DashLine(410,-106)(430,-106){3}
\Line(430,-104)(460,-104)
\Line(430,-106)(460,-106)
\ArrowLine(400,-75.5)(410,-104.5)
\ArrowLine(430,-75.5)(460,-75.5)
\COval(380,-90)(16.5,10)(0){Black}{Red}
\COval(430,-90)(16.5,10)(0){Black}{Red}
\CBoxc(400,-75.5)(5,5){Green}{Green}
\PText(380,-90)(0)[c]{PC}
\PText(430,-90)(0)[c]{PC}

\Line(190,-140)(200,-140)
\Line(195,-145)(195,-135)

\Line(210,-124.5)(240,-124.5)
\Line(210,-126.5)(240,-126.5)
\DashLine(240,-124.5)(260,-124.5){3}
\DashLine(240,-126.5)(260,-126.5){3}
\ArrowLine(260,-125.5)(290,-125.5)
\ArrowLine(210,-155)(240,-155)
\ArrowLine(240,-155)(270,-155)
\Line(270,-154)(320,-154)
\Line(270,-156)(320,-156)
\ArrowLine(260,-125.5)(270,-154.5)
\ArrowLine(290,-125.5)(320,-125.5)
\COval(240,-140)(16.5,10)(0){Black}{Red}
\COval(290,-140)(16.5,10)(0){Black}{Red}
\CBoxc(260,-125.5)(5,5){Green}{Green}
\PText(240,-140)(0)[c]{PC}
\PText(290,-140)(0)[c]{PC}

\Line(330,-140)(340,-140)
\Line(335,-145)(335,-135)

\DashLine(380,-124.5)(400,-124.5){3}
\DashLine(380,-126.5)(400,-126.5){3}
\Line(350,-124.5)(380,-124.5)
\Line(350,-126.5)(380,-126.5)
\ArrowLine(400,-125.5)(430,-125.5)
\ArrowLine(350,-155)(380,-155)
\ArrowLine(380,-155)(410,-155)
\DashLine(410,-154)(430,-154){3}
\DashLine(410,-156)(430,-156){3}
\Line(430,-154)(460,-154)
\Line(430,-156)(460,-156)
\ArrowLine(400,-125.5)(410,-154.5)
\ArrowLine(430,-125.5)(460,-125.5)
\COval(380,-140)(16.5,10)(0){Black}{Red}
\COval(430,-140)(16.5,10)(0){Black}{Red}
\CBoxc(400,-125.5)(5,5){Green}{Green}
\PText(380,-140)(0)[c]{PC}
\PText(430,-140)(0)[c]{PC}

\end{picture}\end{center}
\vspace{.4cm}
\caption{Parity-Violating Diagrams at LO (Note diagrams where lower vertices are parity-violating are not included)}
\label{fig:LO_Diagrams}
\end{figure}
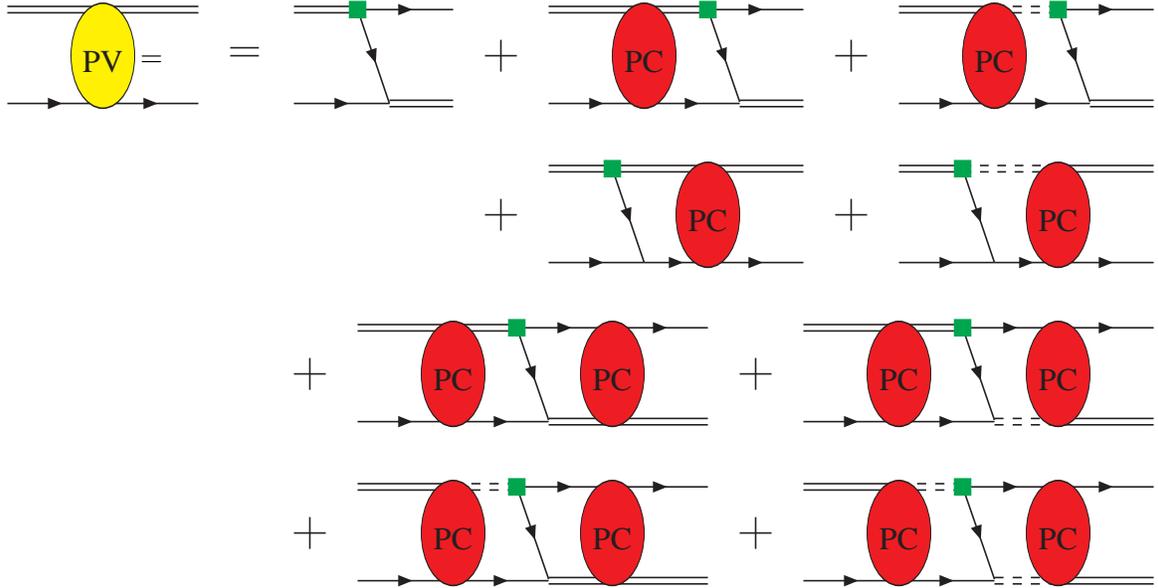

 \noindent using a sharp cutoff $\Lambda$.  The three-body force term is cutoff dependent.  This cutoff is convenient because it can be implemented straightforwardly numerically.  The three-body force term in the Doublet S wave channel is given by\cite{Bedaque:1998kg,Bedaque:1998km}
 
\begin{equation}
{\mathcal{H}}(E,\Lambda)=\frac{2H_{0}(\Lambda)}{\Lambda^{2}}
\end{equation}

\noindent The cutoff dependence of three-body force term $H_{0}(\Lambda)$ is chosen so that the Doublet S wave amplitude produces the correct scattering length\cite{Griesshammer:2004pe,Bedaque:2002yg,Bedaque:1999ve}.  (As noted earlier there is no need to include a parity-violating three-body force, as it has been explicitly shown that no such term exists up to and including NLO\cite{Griesshammer:2010nd}.)  



\indent To first order in parity-violation the integral equation for the parity-violating amplitude is given by the sum of diagrams shown in Fig. \ref{fig:LO_Diagrams}.  The first diagram corresponds to a pure parity-violating transition with no scattering in the initial or final channel.  The next set of diagrams has a parity-violating transition with scattering either in the initial or final channel but not both.  (Also note that for these diagrams the singlet field acts as an intermediate state, which can only exist in the Doublet channel.)  Finally we have the set of diagrams with a parity-violating transition and scattering in both the final and initial channels.\newline\indent Summing all of these figures one finds the parity-violating amplitude given in Eq. (\ref{eq:LO_Amp}), where $\vect{k}$ is the incoming nucleon momentum and $\vect{p}$ is the outgoing nucleon momentum in the c.m. frame.  Since our diagrams are on shell we have $|\vect{k}|=|\vect{p}|$, and the total energy in the c.m. frame is given by $E=\frac{3\vect{k}^{2}}{4M_{N}}-\frac{\gamma_{d}^{2}}{M_{N}}$. The vector index letter $i$ ($j$), represents  the initial (final) deuteron auxiliary field polarization.  Finally the Greek index $\alpha$ ($\beta$) is the initial (final) spinor index and $a$ ($b$) is the initial (final) isospinor index.

\begin{align}
\label{eq:LO_Amp}
\left(it^{ji}_{PV}\right)^{\beta b}_{\alpha a}(\vect{k},\vect{p})&=\frac{4M_{N}}{\sqrt{8}}\frac{i}{\vect{k}^{2}+\vect{k}\cdot\vect{p}+\vect{p}^{2}-M_{N}E-i\epsilon}\left(\mathcal{K}^{11}_{PV}{}^{ji}\right)^{\beta b}_{\alpha a}(\vect{p},\vect{k})\\\nonumber
&+\frac{4M_{N}}{\sqrt{8}}\int\frac{d^{4}q}{(2\pi)^{4}}\mathbf{v}_{p}^{T}(i\tilde{\boldsymbol{\mathcal{K}}}^{jk})^{\beta b}_{\gamma c}(\vect{q},\vect{p},q_{0})i\mathbf{D}\left(\frac{\vect{k}^{2}}{4M_{N}}-\frac{\gamma_{d}^{2}}{M_{N}}+q_{0},\vect{q}\right)\left(\left(it^{ki}\right)^{\gamma c}_{\alpha a}(\vect{k},\vect{q})\right)\times\\\nonumber
&\hspace{1.5cm}\times\propagator{\frac{\vect{k}^{2}}{2M_{N}}-q_{0}}{q}\\\nonumber
&+\frac{4M_{N}}{\sqrt{8}}\int\frac{d^{4}q}{(2\pi)^{4}}\left(\left(it^{jk}\right)^{\beta b}_{\gamma c}(\vect{p},\vect{q})\right)^{T}i\mathbf{D}\left(\frac{\vect{k}^{2}}{4M_{N}}-\frac{\gamma_{d}^{2}}{M_{N}}+q_{0},\vect{q}\right)(i\tilde{\boldsymbol{\mathcal{K}}}^{ki})^{\gamma c}_{\alpha a}(\vect{k},\vect{q},q_{0})\mathbf{v}_{p}\times\\\nonumber
&\hspace{1.5cm}\times\propagator{\frac{\vect{k}^{2}}{2M_{N}}-q_{0}}{q}\\\nonumber
&+\frac{4M_{N}}{\sqrt{8}}\int\frac{d^{4}q}{(2\pi)^{4}}\int\frac{d^{4}\ell}{(2\pi)^{4}}\left(\left(it^{jl}\right)^{\beta b}_{\delta d}(\vect{p},\vectS{\ell})\right)^{T}i\mathbf{D}\left(\frac{\vect{k}^{2}}{4M_{N}}-\frac{\gamma_{d}^{2}}{M_{N}}+q_{0},\vect{q}\right)\\\nonumber
&\hspace{1.5cm}(i\tilde{\boldsymbol{\mathcal{K}}}^{lk})^{\delta d}_{\gamma c}(\vect{q},\vectS{\ell},q_{0}+\ell_{0})i\mathbf{D}\left(\frac{\vect{k}^{2}}{4M_{N}}-\frac{\gamma_{d}^{2}}{M_{N}}+\ell_{0},\vectS{\ell}\right)\left(\left(it^{ki}\right)^{\gamma c}_{\alpha a}(\vect{k},\vect{q})\right)\times\\\nonumber
&\hspace{1.5cm}\times\frac{i}{\frac{\vect{k}^{2}}{2M_{N}}-q_{0}-\frac{\vect{q}^{2}}{2M_{N}}+i\epsilon}\frac{i}{\frac{\vect{k}^{2}}{2M_{N}}-\ell_{0}-\frac{\vectS{\ell}^{2}}{2M_{N}}+i\epsilon}\\\nonumber
\end{align}

\noindent The vector $\mathbf{v}_{p}$ projects out the nucleon-deuteron amplitude in cluster-configuration space and is defined as\cite{Griesshammer:2004pe}

\begin{equation}
\mathbf{v}_{p}=\left(
\begin{array}{c}
1\\
0
\end{array}
\right)
\end{equation}

\noindent and the parity-conserving amplitudes $t$ are a vector defined as follows 

\begin{equation}
\left(\left(it^{ji}\right)^{\beta b}_{\alpha a}(\vect{k},\vect{q})\right)=\left(
\begin{array}{c}
\left(it^{ji}_{Nd\to Nd}\right)^{\beta b}_{\alpha a}(\vect{k},\vect{q})\\
\left(it^{ji}_{Nd\to Nt}\right)^{\beta b}_{\alpha a}(\vect{k},\vect{q})
\end{array}\right)
\end{equation}

\noindent where $t_{Nd\to Nd}$ is the amplitude for $nd$ scattering and $t_{Nd\to Nt}$ is the amplitude for $nd$ going to a nucleon and a singlet combination of the remaining nucleons.  (Note that we have not yet projected out Quartet or Doublet channels.)  The expressions $\mathbf{D}(E,\vect{q})$ and $(i\tilde{\boldsymbol{\mathcal{K}}}^{ji})^{\beta b}_{\alpha a}(\vect{q},\vectS{\ell},q_{0})$ are both matrices defined via.

\begin{equation}
i\mathbf{D}(E,\vect{q})=
\left(
\begin{array}{cc}
iD_{d}(E,\vect{q}) & 0 \\
0&iD_{t}(E,\vect{q})  
\end{array}\right)
\end{equation}

\begin{align}
\label{eq:K}
(i\tilde{\boldsymbol{\mathcal{K}}}^{ji})^{\beta b}_{\alpha a}(\vect{q},\vectS{\ell},q_{0})=&\frac{i}{\frac{1}{2}\vect{q}^{2}+\vect{q}\cdot\vectS{\ell}+\frac{1}{2}\vectS{\ell}^{2}+\frac{1}{4}\vect{k}^{2}+\gamma_{d}^{2}-M_{N}q_{0}-i\epsilon}\times\\\nonumber
&\quad\quad\quad\times\left(
\begin{array}{cc}
\left(\mathcal{K}_{PV}^{11}{}^{ji}\right)^{\beta b}_{\alpha a}(\vect{q},\vectS{\ell}) & \left(\mathcal{K}_{PV}^{12}{}^{ji}\right)^{\beta b}_{\alpha a}(\vect{q},\vectS{\ell})\\
\left(\mathcal{K}_{PV}^{21}{}^{ji}\right)^{\beta b}_{\alpha a}(\vect{q},\vectS{\ell}) & \left(\mathcal{K}_{PV}^{22}{}^{ji}\right)^{\beta b}_{\alpha a}(\vect{q},\vectS{\ell})
\end{array}\right)
\end{align}

\noindent where the functions $\left(\mathcal{K}_{PV}^{XY}{}^{ji}\right)^{\beta b}_{\alpha a}(\vect{q},\vectS{\ell})$, which contain all of the parity-violating dependence are defined as

\begin{subequations}
\label{seq:Kequations}
\begin{align}
\label{eq:K11}
\left(\mathcal{K}_{PV}^{11}{}^{ji}\right)^{\beta b}_{\alpha a}(\vect{k},\vect{p})&=y_{d}\gThreeSOneOnePOne(\sigma^{j})^{\beta}_{\alpha}\delta^{b}_{a}(\vect{k}+2\vect{p})^{i}+iy_{d}\gThreeSOneThreePOne\epsilon^{i\ell k}(\sigma^{k}\sigma^{j})^{\beta}_{\alpha}(\tau_{3})^{b}_{a}(\vect{k}+2\vect{p})^{\ell}\\\nonumber
&+y_{d}\gThreeSOneOnePOne(\sigma^{i})^{\beta}_{\alpha}\delta^{b}_{a}(2\vect{k}+\vect{p})^{j}-iy_{d}\gThreeSOneThreePOne\epsilon^{j\ell k}(\sigma^{i}\sigma^{k})^{\beta}_{\alpha}(\tau_{3})^{b}_{a}(2\vect{k}+\vect{p})^{\ell}
\end{align}
\vspace{-2cm}

\begin{align}
\left(\mathcal{K}_{PV}^{12}{}^{jA}\right)^{\beta b}_{\alpha a}(\vect{k},\vect{p})&=y_{d}\gDeltaIZero(\sigma^{\ell}\sigma^{j})^{\beta}_{\alpha}(\tau^{A})^{b}_{a}(\vect{k}+2\vect{p})^{\ell}+iy_{d}\gDeltaIOne\epsilon^{3AC}(\sigma^{\ell}\sigma^{j})^{\beta}_{\alpha}(\tau^{C})^{b}_{a}(\vect{k}+2\vect{p})^{\ell}\\\nonumber
&+y_{t}\gThreeSOneOnePOne\delta^{\beta}_{\alpha}(\tau^{A})^{b}_{a}(2\vect{k}+\vect{p})^{j}-iy_{t}\gThreeSOneThreePOne\epsilon^{j\ell k}(\tau^{A}\tau_{3})^{b}_{a}(\sigma^{k})^{\beta}_{\alpha}(2\vect{k}+\vect{p})^{\ell}
\end{align}
\vspace{-2cm}

\begin{align}
\left(\mathcal{K}_{PV}^{21}{}^{Bi}\right)^{\beta b}_{\alpha a}(\vect{k},\vect{p})&=y_{t}\gThreeSOneOnePOne(\tau^{B})^{b}_{a}\delta^{\beta}_{\alpha}(\vect{k}+2\vect{p})^{i}+iy_{t}\gThreeSOneThreePOne\epsilon^{i\ell k}(\sigma^{k})^{\beta}_{\alpha}(\tau_{3}\tau^{B})^{b}_{a}(\vect{k}+2\vect{p})^{\ell}\\\nonumber
&+y_{d}\gDeltaIZero(\sigma^{i}\sigma^{\ell})^{\beta}_{\alpha}(\tau^{B})^{b}_{a}(2\vect{k}+\vect{p})^{\ell}-iy_{d}\gDeltaIOne\epsilon^{3BC}(\sigma^{i}\sigma^{\ell})^{\beta}_{b}(\tau^{C})^{b}_{a}(2\vect{k}+\vect{p})^{\ell}
\end{align}
\vspace{-2cm}

\begin{align}
\left(\mathcal{K}_{PV}^{22}{}^{BA}\right)^{\beta b}_{\alpha a}(\vect{k},\vect{p})&=y_{t}\gDeltaIZero(\sigma^{\ell})^{\beta}_{\alpha}(\tau^{A}\tau^{B})^{b}_{a}(\vect{k}+2\vect{p})^{\ell}+iy_{t}\gDeltaIOne\epsilon^{3AC}(\sigma^{\ell})^{\beta}_{\alpha}(\tau^{C}\tau^{B})^{b}_{a}(\vect{k}+2\vect{p})^{\ell}\\\nonumber
&+y_{t}\gDeltaIZero(\tau^{A}\tau^{B})^{b}_{a}(\sigma^{\ell})^{\beta}_{\alpha}(2\vect{k}+\vect{p})^{\ell}-iy_{t}\gDeltaIOne\epsilon^{3BC}(\tau^{A}\tau^{C})^{b}_{a}(\sigma^{\ell})^{\beta}_{\alpha}(2\vect{k}+\vect{p})^{\ell}
\end{align}
\end{subequations}

\noindent (Note that the capital letters A,B, and C are used for the singlet auxiliary field polarization and the lowercase letters i,j, and k are used for the deuteron auxiliary field polarization.)  Integrating over the energy and picking up the poles from the nucleon propagators in our diagrams Eq. (\ref{eq:LO_Amp}) becomes.

\begin{align}
\left(t^{ji}_{PV}\right)^{\beta b}_{\alpha a}(\vect{k},\vect{p})&=\frac{4M_{N}}{\sqrt{8}}\mathbf{v}_{p}^{T}\left(\boldsymbol{\mathcal{K}}^{ji}\right)^{\beta b}_{\alpha a}(\vect{k},\vect{p})\mathbf{v}_{p}\\\nonumber
&-\frac{4M_{N}}{\sqrt{8}}\int\frac{d^{3}q}{(2\pi)^{3}}\mathbf{v}_{p}^{T}(\boldsymbol{\mathcal{K}}^{jk})^{\beta b}_{\gamma c}(\vect{q},\vect{p})\mathbf{D}\left(E-\frac{\vect{q}^{2}}{2M_{N}},\vect{q}\right)\left(\left(t^{ki}\right)^{\gamma c}_{\alpha a}(\vect{k},\vect{q})\right)\\\nonumber
&-\frac{4M_{N}}{\sqrt{8}}\int\frac{d^{3}q}{(2\pi)^{3}}\left(\left(t^{jk}\right)^{\beta b}_{\gamma c}(\vect{q},\vect{p})\right)^{T}\mathbf{D}\left(E-\frac{\vect{q}^{2}}{2M_{N}},\vect{q}\right)(\boldsymbol{\mathcal{K}}^{ki})^{\gamma c}_{\alpha a}(\vect{k},\vect{q})\mathbf{v}_{p}\\\nonumber
&+\frac{4M_{N}}{\sqrt{8}}\int\frac{d^{3}q}{(2\pi)^{3}}\int\frac{d^{3}\ell}{(2\pi)^{3}}\left(\left(t^{jl}\right)^{\beta b}_{\delta d}(\vectS{\ell},\vect{p},)\right)^{T}\mathbf{D}\left(E-\frac{\vect{q}^{2}}{2M_{N}},\vect{q}\right)\\\nonumber
&\hspace{1.5cm}(\boldsymbol{\mathcal{K}}^{lk})^{\delta d}_{\gamma c}(\vect{q},\vectS{\ell})\mathbf{D}\left(E-\frac{\vectS{\ell}^{2}}{2M_{N}},\vectS{\ell}\right)\left(\left(t^{ki}\right)^{\gamma c}_{\alpha a}(\vect{k},\vect{q})\right)
\end{align}

\noindent where 

\begin{align}
\label{eq:Krelation}
(\boldsymbol{\mathcal{K}}^{ji})^{\beta b}_{\alpha a}(\vect{q},\vectS{\ell})=&\frac{1}{\vect{q}^{2}+\vect{q}\cdot\vectS{\ell}+\vectS{\ell}^{2}-M_{N}E-i\epsilon}\times\\\nonumber
&\quad\quad\quad\times\left(
\begin{array}{cc}
\left(\mathcal{K}_{PV}^{11}{}^{ji}\right)^{\beta b}_{\alpha a}(\vect{q},\vectS{\ell}) & \left(\mathcal{K}_{PV}^{12}{}^{ji}\right)^{\beta b}_{\alpha a}(\vect{q},\vectS{\ell})\\
\left(\mathcal{K}_{PV}^{21}{}^{ji}\right)^{\beta b}_{\alpha a}(\vect{q},\vectS{\ell}) & \left(\mathcal{K}_{PV}^{22}{}^{ji}\right)^{\beta b}_{\alpha a}(\vect{q},\vectS{\ell})
\end{array}\right)
\end{align}

 Now that we have derived the parity-violating amplitude, we note that it contains the related scattering amplitudes from the parity-conserving sector.  Such parity-conserving scattering amplitudes are calculated in\cite{Griesshammer:2004pe}, by numerically solving Fadeev's equation in an angular momentum basis.  However, as part of this solution one runs into singularities along the real axis.  To overcome this difficulty the method of Hetherington and Schick is employed, in which the axis of integration is rotated into the complex plane, therefore avoiding the singularities\cite{Hetherington:1965zza,Cahill:1971,Aaron:1966zz}.  One can then use the solutions along the deformed contour to solve for the amplitudes along the real axis. Details of the procedure to calculate these amplitudes can be found in\cite{Ziegelmann}.  In order to use the solutions to Fadeev's equations we need to project out our parity-violating amplitude into an angular momentum basis.  However, unlike the parity-conserving sector, the PV amplitudes mix different angular momentum states. Also, since at leading order, spin and orbital angular momentum mix, the appropriate angular momentum basis to use is the total angular momentum $\vect{J}=\vect{L}+\vect{S}$.  Thus we express our parity-violating amplitude as 

\begin{equation}
t_{PV}(\vect{k},\vect{p})=\sum_{J=0}^{\infty}\sum_{M=-J}^{M=J}\sum_{L=|J-S|}^{J+S}\sum_{L'=|J-S'|}^{J+S'}\sum_{S,S'}4\pi t^{JM}_{L'S',LS}(k,p)\mathscr{Y}^{M}_{J,L'S'}(\hat{\mathbf{p}})\left(\mathscr{Y}^{M}_{J,LS}(\hat{\mathbf{k}})\right)^{*}
\end{equation}

\noindent where the spin angle functions are given by

\begin{equation}
\mathscr{Y}^{M}_{J,LS}(\hat{\mathbf{k}})=\sum_{m_{L},m_{S}}C_{L,S;J}^{m_{L},m_{S},M}Y_{L}^{m_{L}}(\hat{\mathbf{k}})\chi_{S}^{m_{S}}
\end{equation}

\noindent $\chi_{S}^{m_{S}}$ being the spinor part of the spin-angle functions, $C_{L,S,J}^{m_{L},m_{S},M}$ the appropriate Clebsch-Gordan coefficient, and the appropriate $Y_{L}^{m_{L}}(\hat{\mathbf{k}})$ spherical harmonic.  Since the spin-angle functions are orthogonal, we can project out the amplitudes in our angular momentum basis via

\begin{equation}
 t^{JM}_{L'S',LS}(k,p)=\frac{1}{4\pi}\int d\Omega_{k}\int d\Omega_{p}\left(\mathscr{Y}^{M}_{J,L'S'}(\hat{\mathbf{p}})\right)^{*} t_{PV}(\vect{k},\vect{p})\mathscr{Y}^{M}_{J,LS}(\hat{\mathbf{k}})
\end{equation}

At sufficiently low energies S-P mixing will dominate.  Thus we will calculate only the amplitudes $t^{SM}_{1S',0S}$  (Note J=S here since L=0) for all possible values of $S$ and $S'$.  All spin and angle dependence is contained within the matrix $(\boldsymbol{\mathcal{K}}^{ij})^{\beta b}_{\alpha a}(\vect{q},\vectS{\ell})$, and the appropriate projections in $\vect{J}$, $\vect{L}$,and $\vect{S}$ can be found in the appendix.  Going to a partial wave basis we finally obtain an expression for the parity-violating partial wave amplitudes.

\begin{align}
\label{eq:LO_Amp_Projected}
{t_{PV}}^{JM}_{L'S',LS}(k,p)&=\frac{M_{N}}{\sqrt{8}\pi}\mathbf{v}_{p}^{T}\boldsymbol{\mathcal{K}}(k,p)^{J}_{L'S',LS}\mathbf{v}_{p}+\\\nonumber
&-\frac{M_{N}}{2\sqrt{8}\pi^{3}}\int_{0}^{\infty}dqq^{2}\mathbf{v}_{p}^{T}\boldsymbol{\mathcal{K}}(q,p)^{J}_{L'S',LS}\mathbf{D}\left(E-\frac{\vect{q}^{2}}{2M_{N}},\vect{q}\right)\left({t_{PC}}^{JM}_{LS,LS}(k,q)\right)\\\nonumber
&-\frac{M_{N}}{2\sqrt{8}\pi^{3}}\int_{0}^{\infty}dqq^{2}\left({t_{PC}}^{JM}_{L'S',L'S'}(q,p)\right)^{T}\mathbf{D}\left(E-\frac{\vect{q}^{2}}{2M_{N}},\vect{q}\right)\boldsymbol{\mathcal{K}}(k,q)^{J}_{L'S',LS}\mathbf{v}_{p}\\\nonumber
&+\frac{M_{N}}{4\sqrt{8}\pi^{5}}\int_{0}^{\infty}dq q^{2}\int_{0}^{\infty}d\ell\ell^{2}\left({t_{PC}}^{JM}_{L'S',L'S'}(p,\ell)\right)^{T}\mathbf{D}\left(E-\frac{\vect{q}^{2}}{2M_{N}},\vect{q}\right)\\\nonumber
&\quad\quad\boldsymbol{\mathcal{K}}(q,\ell)^{JM}_{L'S',LS}\mathbf{D}\left(E-\frac{\vectS{\ell^{2}}}{2M_{N}},\vectS{\ell}\right)\left({t_{PC}}^{JM}_{LS,LS}(k,q)\right)
\end{align}

\noindent This expression contains the parity-conserving amplitudes in the partial wave basis of total angular momentum $\vect{J}=\vect{L}+\vect{S}$.  (These are equivalent to the parity-conserving amplitudes in the partial wave basis of orbital angular momentum.)  It can be shown straightforwardly that the parity-conserving amplitudes are independent of total angular momentum $\vect{J}$.  Thus we can use the parity-conserving amplitudes as calculated numerically by\cite{Gabbiani:1999yv,Griesshammer:2004pe,Bedaque:2002yg} and perform the integration numerically in order to obtain the associated parity-violating amplitudes.

Before integrating Eq. (\ref{eq:LO_Amp_Projected}) we multiply by the LO deuteron renormalization $Z_{D}=(8\pi\gamma_{d})/(M_{N}^{2}y_{d}^{2})$\cite{Gabbiani:1999yv}, and use the renormalized parity-conserving amplitudes.  We find that all the parity-violating LEC's occur in the combinations.  

\begin{displaymath}
\frac{g^{{}^{3}\!S_{1}-{}^{1}\!P_{1}}}{y_{d}},\frac{g^{{}^{3}\!S_{1}-{}^{3}\!P_{1}}}{y_{d}},\frac{g^{{}^{1}\!S_{0}-{}^{3}\!P_{0}}_{(\Delta I=0)}}{y_{t}},\frac{g^{{}^{1}\!S_{0}-{}^{3}\!P_{0}}_{(\Delta I=1)}}{y_{t}}
\end{displaymath}

\noindent (Note $g^{{}^{1}\!S_{0}-{}^{3}\!P_{0}}_{(\Delta I=2)}$ does not appear as a $\Delta I=2$ transition is not allowed for a first order parity-violating transition in $nd$ scattering.)  For the sake of convenience we find it useful to make the following definitions.

\begin{displaymath}
g_{1}=\frac{g^{{}^{3}\!S_{1}-{}^{1}\!P_{1}}}{y_{d}},g_{2}=\frac{g^{{}^{3}\!S_{1}-{}^{3}\!P_{1}}}{y_{d}},g_{3}=\frac{g^{{}^{1}\!S_{0}-{}^{3}\!P_{0}}_{(\Delta I=0)}}{y_{t}},g_{4}=\frac{g^{{}^{1}\!S_{0}-{}^{3}\!P_{0}}_{(\Delta I=1)}}{y_{t}},g_{5}=\frac{g^{{}^{1}\!S_{0}-{}^{3}\!P_{0}}_{(\Delta I=2)}}{y_{t}}
\end{displaymath}

\noindent Since these coefficients are unknown, we will write the parity-violating partial wave amplitudes as follows, where $\left({t_{PV}}^{JM}_{L'S',LS}(k,p)\right)^{i}$ is calculated by setting $g_{j}=0, j\neq i$ and $g_{i}=1$ in the parity-violating partial wave amplitude.  Thus the parity-violating amplitude can be written as.

\begin{equation}
\label{eq:PVAmp}
{t_{PV}}^{JM}_{L'S',LS}(k,p)=\sum_{i=1}^{4}g_{i}
\left({t_{PV}}^{JM}_{L'S',LS}(k,p)\right)^{i}
\end{equation}

\vspace{.2cm}
\section{Parity-Violating Potential}
It is clear from Eq. (\ref{eq:PVAmp}) that in order to obtain numerical values for the parity-violating amplitude one needs to know the size of the coefficients $g_{i}$, which at this time are not determined from either theory or experiment.  Nevertheless, we can obtain estimates by matching the $g_{i}$ onto the familiar DDH coefficients.  We will carry out this procedure in three steps.  First we match the DDH coefficients onto the coefficients of the Zhu potential\cite{Zhu:2004vw}.  Then we match the Zhu potential on to the Girlanda potential\cite{Schiavilla:2008ic,PhysRevC.83.029902}.  Finally we project the coefficients of the Girlanda potential onto the coefficients of the auxiliary field formalism.  We also show how all these formalisms can be matched to the familiar Danilov parameters\newline\indent The DDH model\cite{Desplanques:1979hn} is a single-meson-exchange picture, limited to exchange of the lightest mesons $\pi$, $\rho$, and $\omega$.\footnote{Since CP is conserved there are no neutral pseudoscalar mesons $\pi^{0}$, $\eta$, or $\eta'$ by Barton's theorem\cite{Barton:1961eg}}  The strong Hamiltonian is given by

\begin{align}
{\mathcal{H}}_{st}=&ig_{\pi NN}\bar{N}\gamma_{5}\tau\cdot\pi N+g_{\rho}\bar{N}\left(\gamma_{\mu}+i\frac{(1+\chi_{\rho})}{2M_{N}}\sigma_{\mu\nu}k^{\nu}\right)\tau\cdot\rho^{\mu}N\\\nonumber
&+g_{\omega}\bar{N}\left(\gamma_{\mu}+i\frac{(1+\chi_{\omega})}{2M_{N}}\sigma_{\mu\nu}k^{\nu}\right)\omega^{\mu}N
\end{align}

\noindent with the strong couplings given approximately by $g_{\pi NN}^{2}/4\pi\simeq 13.5$ and $g_{\rho}^{2}/4\pi=\frac{1}{9}g_{\omega}^{2}/4\pi\simeq.67$, while the magnetic moment terms are approximately $\chi_{\rho}=\kappa_{p}-\kappa_{n}=3.7$ and $\chi_{\omega}=\kappa_{p}+\kappa_{n}=-.12$.  The phenomenological weak interaction Hamiltonian posited by DDH consists of seven weak coupling terms

\begin{align}
{\mathcal{H}}_{wk}=&i\frac{f_{\pi}^{1}}{\sqrt{2}}\bar{N}(\tau\times\pi)_{z}N+\bar{N}\left(h_{\rho}^{0}\tau\cdot\rho^{\mu}+h_{\rho}^{1}\rho_{z}^{\mu}+\frac{g_{\rho}^{2}}{2\sqrt{6}}(3\tau_{z}\rho_{z}^{\mu}-\tau\cdot\rho^{\mu})\right)\gamma_{\mu}\gamma_{5}N\\\nonumber
&+\bar{N}\left(h_{\omega}^{0}\omega^{\mu}+h_{\omega}^{1}\tau_{z}\omega^{\mu}\right)\gamma_{\mu}\gamma_{5}N-{h_{\rho}'}^{1}\bar{N}(\tau\times\rho^{\mu})_{z}\frac{\sigma_{\mu\nu}k^{\nu}}{2M_{N}}\gamma_{5}N
 \end{align}
 
\noindent DDH attempted to obtain theoretical predictions for the seven constants using SU(6) symmetry and quark model techniques.  However, due to the difficulty of this calculation they were only able to come up with reasonable ranges and ``best" values as shown in Table \ref{tab:DDH}.  (Also shown are estimates by other groups.)

\begin{table}[hbt]
\begin{center}
\begin{tabular}{|c|c|c|c|c|}
\hline
{} & DDH\cite{Desplanques:1979hn} & DDH\cite{Desplanques:1979hn} & DZ\cite{Dubovik:1986pj} & FCDH\cite{Feldman:1991tj}\\
Coupling & Reasonable Range & ``Best" Value &{} & {}\\\hline
$f_{\pi}$ & $0\to 30$ & +12 & +3 & +7\\
$h^{0}_{\rho}$ & $30\to-81$ & -30 & -22 & -10\\
$h^{1}_{\rho}$ & $-1\to 0$ & -.5 & +1 & -1\\
$h^{2}_{\rho}$ & $-20\to -29$ & -25 & -18 & -18\\
$h^{0}_{\omega}$ & $15\to -27$ & -5 & -10 & -13\\
$h^{1}_{\omega}$ & $-5\to -2$ & -3 & -6 & -6\\\hline
\end{tabular}
\end{center}
\caption{\label{tab:DDH}Weak NNM couplings.  All numbers are quoted in units of the "sum rule" value\newline $S_{R}=3.8\times 10^{-8}$}
\end{table}

The form of any parity-violating potential can be written as a sum of operators $O_{ij}^{(n)}$ with corresponding coefficients $c_{n}^{\alpha}$, where $\alpha$ refers to the specific potential of interest.. 

\begin{equation}
\label{eq:PVpotential}
V_{ij}^{\alpha}=\sum_{n}c_{n}^{\mathrm{\alpha}}O_{ij}^{(n)}
\end{equation}

\noindent At the lowest energies the component of the operators that contain momentum is of two forms.

\vspace{-.5cm}

\begin{align}
\mathbf{X}^{(n)}_{ij,+}&=\{\vect{p}_{ij},f_{n}^{\alpha}(r_{ij})\}\\\nonumber
\mathbf{X}^{(n)}_{ij,-}&=i[\vect{p}_{ij},f_{n}^{\alpha}(r_{ij})]\\\nonumber
\end{align}

\noindent where $\vect{p}_{ij}=(\vect{p}_{1}-\vect{p}_{2})/2$ is the momentum of the nucleon-nucleon system in the c.m. frame. 

\begin{table}[h]
\caption{\label{tab:DDHpotential}Parity-violating potential in DDH and Zhu formalism.  ${\cal T}_{ij}\equiv (3\tau_i^z\tau_j^z-\tau_i\cdot\tau_j)$. (Note $\Lambda_{\chi}^{3}\sim 4\pi F_{\pi}$ is the chiral scale\cite{Manohar:1983md,PhysRevD:30:587}, where $F_{\pi}=92.4 MeV$ is the pion decay constant)}
\begin{ruledtabular}
\begin{center}
\begin{tabular}{cccccc}
$n$ & $c_n^{DDH}$ & $c_{n}^{Zhu}$ & $f_n^{DDH}(r)$ & $f_{n}^{Zhu}$ &  $O^{(n)}_{ij}$ \\
\hline
$1$ & $+\frac{g_{\pi NN}}{2\sqrt{2} M_N}f_\pi$ & $\frac{1}{\Lambda_{\chi}^{3}}2\tilde{C}_{6}$ & $f_\pi(r)$ & $f_{m}(r)$ & $(\tau_i\times\tau_j)^z(\vectS{\sigma}_i+\vectS{\sigma}_j)\cdot{\boldsymbol{X}}^{(1)}_{ij,-}$
\\
$2 $ & $ -\frac{g_\rho}{M_N}h_\rho^0 $ & $\frac{1}{\Lambda_{\chi}^{3}}2C_{3}$ & $ f_\rho(r) $ &  $f_{m}(r)$ & 
       $(\tau_i\cdot\tau_j)(\vectS{\sigma}_i-\vectS{\sigma}_j)\cdot{\boldsymbol{X}}^{(2)}_{ij,+}$
\\
$3 $ & $ -\frac{g_\rho(1+\chi_\rho)}{M_N} h_\rho^0 $ & $\frac{1}{\Lambda_{\chi}^{3}}2\tilde{C}_{3}$ & $f_\rho(r) $ & $f_{m}(r)$ & 
      $(\tau_i\cdot\tau_j)(\vectS{\sigma}_i\times\vectS{\sigma}_j)\cdot{\boldsymbol{X}}^{(3)}_{ij,-}$
\\
$4 $ & $ -\frac{g_\rho}{2 M_N} h_\rho^1 $ & $\frac{1}{\Lambda_{\chi}^{3}}C_{4}$ & $ f_\rho(r) $ & $f_{m}(r)$ &  
      $(\tau_i+\tau_j)^z(\vectS{\sigma}_i-\vectS{\sigma}_j)\cdot{\boldsymbol{X}}^{(4)}_{ij,+}$
\\
$5  $ & $  -\frac{g_\rho(1+\chi_\rho)}{2 M_N}h_\rho^1 $ & $\frac{1}{\Lambda_{\chi}^{3}}\tilde{C}_{4}$ & $ f_\rho(r)
     $ & $f_{m}(r)$ & 
      $(\tau_i+\tau_j)^z(\vectS{\sigma}_i\times\vectS{\sigma}_j)\cdot{\boldsymbol{X}}^{(5)}_{ij,-}$
\\
$6 $ & $ -\frac{g_\rho}{2\sqrt{6} M_N}h_\rho^2 $ & $\frac{1}{\Lambda_{\chi}^{3}}2C_{5}$ & $ f_\rho(r) $  & $f_{m}(r)$ &  
  ${\cal T}_{ij}
   (\vectS{\sigma}_i-\vectS{\sigma}_j)\cdot{\boldsymbol{X}}^{(6)}_{ij,+}$
\\
$7 $ & $ -\frac{g_\rho(1+\chi_\rho)}{2\sqrt{6} M_N}h_\rho^2 $ & $\frac{1}{\Lambda_{\chi}^{3}}2\tilde{C}_{5}$ & $ f_\rho(r) $ &  $f_{m}(r)$ & ${\cal T}_{ij}(\vectS{\sigma}_i\times\vectS{\sigma}_j)\cdot{\boldsymbol{X}}^{(7)}_{ij,-}$
\\
$8 $ & $ -\frac{g_\omega}{M_N}h_\omega^0 $ & $\frac{1}{\Lambda_{\chi}^{3}}2C_{1}$ & $ f_\omega(r) $ & $f_{m}(r)$ & 
     $(\vectS{\sigma}_i-\vectS{\sigma}_j)\cdot{\boldsymbol{X}}^{(8)}_{ij,+}$
\\
$9  $ & $  -\frac{g_\omega(1+\chi_\omega)}{M_N} h_\omega^0 $ & $\frac{1}{\Lambda_{\chi}^{3}}2\tilde{C}_{1}$ & $ f_\omega(r) $ & $f_{m}(r)$ &
    $  (\vectS{\sigma}_i\times\vectS{\sigma}_j)\cdot{\boldsymbol{X}}^{(9)}_{ij,-}$
\\
$10 $ & $ -\frac{g_\omega}{2 M_N} h_\omega^1 $ & $\frac{1}{\Lambda_{\chi}^{3}}C_{2}$ &$ f_\omega(r) $ & $f_{m}(r)$ &    $(\tau_i+\tau_j)^z(\vectS{\sigma}_i-\vectS{\sigma}_j)\cdot{\boldsymbol{X}}^{(10)}_{ij,+}$
\\
$11 $ & $ -\frac{g_\omega(1+\chi_\omega)}{2M_N} h^1_\omega $ & $\frac{1}{\Lambda_{\chi}^{3}}\tilde{C}_{2}$ & $ f_\omega(r) $ & $f_{m}(r)$  &   $(\tau_i+\tau_j)^z(\vectS{\sigma}_i\times\vectS{\sigma}_j)\cdot{\boldsymbol{X}}^{(11)}_{ij,-}$
\\
$12 $ & $ -\frac{g_\omega h_\omega^1-g_\rho h_\rho^1}{2M_N} $ & $\frac{1}{\Lambda_{\chi}^{3}}(C_{2}-C_{4})$ & $ f_\rho(r) $  &  $f_{m}(r)$ & $(\tau_i-\tau_j)^z(\vectS{\sigma}_i+\vectS{\sigma}_j)\cdot{\boldsymbol{X}}^{(12)}_{ij,+}$
\\
$13 $ & $ -\frac{g_\rho}{2M_N} h^{'1}_\rho $ & $0$ & $ f_\rho(r) $ & $0$ &        $(\tau_i\times\tau_j)^z(\vectS{\sigma}_i+\vectS{\sigma}_j)\cdot{\boldsymbol{X}}^{(13)}_{ij,-}$
\end{tabular}
\end{center}
\end{ruledtabular}
\end{table} 

\noindent The coefficients, operators, and regulator functions $f_{n}^{DDH}(r_{ij})$ for the DDH potential and $f_{n}^{Zhu}(r_{ij})$ for the Zhu potential are given in Table \ref{tab:DDHpotential}. The functions $f_{n}^{DDH}(r_{ij})$ are Yukawa functions, where the mass corresponds to the appropriate meson\cite{Holstein:2010zza}.  However, at the lowest energies the functions for the DDH potential can be written as $f_{i}(r)=\frac{1}{m_{i}^{2}}\delta^{3}(\vect{r})$, where $i=\pi$,$\rho$, or $\omega$\cite{Holstein:2010zza}.  Likewise the functions $f_{m}(r)=\delta^{3}(\vect{r})$ for the Zhu potential become delta functions in the low energy limit.  Thus at low energies the DDH potential and the Zhu potential can be trivially matched yielding\cite{Zhu:2004vw}

\begin{equation}
\frac{\tilde{C}_{1}}{C_{1}}=\frac{\tilde{C}_{2}}{C_{2}}=1+\chi_{\omega}\simeq .88
\end{equation}

\begin{equation}
\frac{\tilde{C}_{3}}{C_{3}}=\frac{\tilde{C}_{4}}{C_{4}}=\frac{\tilde{C}_{5}}{C_{5}}=1+\chi_{\rho}\simeq 4.7
\end{equation}

\begin{align}
C_{1}^{DDH}=&-\frac{\Lambda_{\chi}^{3}}{2M_{N}m_{\omega}^{2}}g_{\omega}h_{\omega}^{0}\stackrel{\mathrm{bestguess}}{\longrightarrow}2.25\times10^{-6}\\\nonumber
C_{2}^{DDH}=&-\frac{\Lambda_{\chi}^{3}}{2M_{N}m_{\omega}^{2}}g_{\omega}h_{\omega}^{1}\stackrel{\mathrm{bestguess}}{\longrightarrow}1.35\times10^{-6}\\\nonumber
C_{3}^{DDH}=&-\frac{\Lambda_{\chi}^{3}}{2M_{N}m_{\rho}^{2}}g_{\rho}h_{\rho}^{0}\stackrel{\mathrm{bestguess}}{\longrightarrow}4.58\times10^{-6}\\\nonumber
C_{4}^{DDH}=&-\frac{\Lambda_{\chi}^{3}}{2M_{N}m_{\rho}^{2}}g_{\rho}h_{\rho}^{1}\stackrel{\mathrm{bestguess}}{\longrightarrow}7.64\times10^{-8}\\\nonumber
C_{5}^{DDH}=&\frac{\Lambda_{\chi}^{3}}{4\sqrt{6}M_{N}m_{\rho}^{2}}g_{\rho}h_{\rho}^{0}\stackrel{\mathrm{bestguess}}{\longrightarrow}-7.80\times10^{-7}\\\nonumber
\tilde{C}_{6}^{DDH}\simeq &\frac{\Lambda_{\chi}^{3}}{4\sqrt{2}M_{N}m_{\pi}^{2}}g_{\pi NN}f_{\pi}\stackrel{\mathrm{bestguess}}{\longrightarrow}9.19\times10^{-5}\\\nonumber
\end{align}

As first pointed out by Danilov, one needs five parity-violating terms at the lowest energies in the two-body sector\cite{Danilov}, since only S-P mixing is involved.  By conservation of angular momentum the state ${}^{3}\!S_{1}$, can only connect with the states ${}^{1}\!P_{1}$ or ${}^{3}\!P_{1}$.  Since ${}^{3}\!S_{1}$ is an isosinglet there is a unique way to get to the isosinglet state ${}^{1}\!P_{1}$ and isotriplet state ${}^{3}\!P_{1}$.  Similarly, the state ${}^{1}\!S_{0}$ can only connect with the state ${}^{3}\!P_{0}$.  However, both ${}^{1}\!S_{0}$ and ${}^{3}\!P_{0}$ are isotriplet states so the operator connecting these states can carry $\Delta I=0,1,$ or $2$.  The existence of five unique operators which characterize parity violation at low energy appears to be in contradiction with the DDH and Zhu potential, which involve ten different operators.  However, at low energies five of these operator structures are redundant as shown by Girlanda\cite{Girlanda:2008ts}.  In this procedure one begins with all possible one-derivative P violating CP conserving relativistic terms.
  
\begin{align}
{\mathcal{O}}_{1}=&\quad\bar{\psi}\gamma^{\mu}\psi\bar{\psi}\gamma_{\mu}\gamma_{5}\psi && \tilde{\mathcal{O}}_{1}=\quad\bar{\psi}\gamma^{\mu}\gamma_{5}\psi\partial^{\nu}(\bar{\psi}\sigma_{\mu\nu}\psi)\\\nonumber
{\mathcal{O}}_{2}=&\quad\bar{\psi}\gamma^{\mu}\psi\bar{\psi}\tau_{3}\gamma_{\mu}\gamma_{5}\psi &&
\tilde{\mathcal{O}}_{2}=\quad\bar{\psi}\gamma^{\mu}\gamma_{5}\psi\partial^{\nu}(\bar{\psi}\tau_{3}\sigma_{\mu\nu}\psi)\\\nonumber
{\mathcal{O}}_{3}=&\quad\bar{\psi}\tau_{a}\gamma^{\mu}\psi\bar{\psi}\tau^{a}\gamma_{\mu}\gamma_{5}\psi &&
\tilde{\mathcal{O}}_{3}=\quad\bar{\psi}\tau_{a}\gamma^{\mu}\gamma_{5}\psi\partial^{\nu}(\bar{\psi}\tau^{a}\sigma_{\mu\nu}\psi)\\\nonumber
{\mathcal{O}}_{4}=&\quad\bar{\psi}\tau_{3}\gamma^{\mu}\psi\bar{\psi}\gamma_{\mu}\gamma_{5}\psi &&
\tilde{\mathcal{O}}_{4}=\quad\bar{\psi}\tau_{3}\gamma^{\mu}\gamma_{5}\psi\partial^{\nu}(\bar{\psi}\sigma_{\mu\nu}\psi)\\\nonumber
{\mathcal{O}}_{5}=&\quad{\mathcal{I}}_{ab}\bar{\psi}\tau_{a}\gamma^{\mu}\psi\bar{\psi}\tau_{b}\gamma_{\mu}\gamma_{5}\psi &&
\tilde{\mathcal{O}}_{5}=\quad{\mathcal{I}}_{ab}\bar{\psi}\tau_{a}\gamma^{\mu}\gamma_{5}\psi\partial^{\nu}(\bar{\psi}\tau_{b}\sigma_{\mu\nu}\psi)\\\nonumber
{\mathcal{O}}_{6}=&\quad i\epsilon_{ab3}\bar{\psi}\tau_{a}\gamma^{\mu}\psi\bar{\psi}\tau_{b}\gamma_{\mu}\gamma_{5}\psi &&
\tilde{\mathcal{O}}_{6}=\quad i\epsilon_{ab3}\bar{\psi}\tau_{a}\gamma^{\mu}\gamma_{5}\psi\partial^{\nu}(\bar{\psi}\tau_{b}\sigma_{\mu\nu}\psi)
\end{align}

Using Fierz transformations and the equations of motion, there exist six identities

\begin{align}
\label{eq:GirlandaResults}
{\mathcal{O}}_{3}=&{\mathcal{O}}_{1}& \tilde{\mathcal{O}}_{2}+\tilde{\mathcal{O}}_{4}=&M_{N}({\mathcal{O}}_{2}+{\mathcal{O}}_{4})\\\nonumber
{\mathcal{O}}_{2}-{\mathcal{O}}_{4}=&2{\mathcal{O}}_{6}& \tilde{\mathcal{O}}_{2}-\tilde{\mathcal{O}}_{4}=&-2M_{N}{\mathcal{O}}_{6}-\tilde{\mathcal{O}}_{6}\\\nonumber
\tilde{\mathcal{O}}_{3}+3\tilde{\mathcal{O}}_{1}=&2M_{N}({\mathcal{O}}_{1}+{\mathcal{O}}_{3})& \tilde{\mathcal{O}}_{5}=&M_{N}{\mathcal{O}}_{5}
\end{align}

\noindent reducing the number of unique operators to six.  However, in a non-relativistic reduction it turns out that two of the operators have equivalent structures leaving five unique operators at the lowest energies.  The resulting parity-violating Lagrangian in the Girlanda formalism is given by

\begin{align}
\label{eq:GirlandaLagrangian}
{\mathcal{L}}_{\mathrm{PV}}^{\mathrm{Gir}}=\ &{\mathcal{G}}_{1}(N^{\dagger}\vectS{\sigma}N\cdot N^{\dagger}i\stackrel{\leftrightarrow}{\nabla}N-N^{\dagger}NN^{\dagger}i\stackrel{\leftrightarrow}{\nabla}\cdot\vectS{\sigma}N)-\tilde{{\mathcal{G}}}_{1}\epsilon_{ijk}N^{\dagger}\sigma_{i}N\nabla_{j}(N^{\dagger}\sigma_{k}N)\\\nonumber
&-{\mathcal{G}}_{2}\epsilon_{ijk}[N^{\dagger}\tau_{3}\sigma_{i}N\nabla_{j}(N^{\dagger}\sigma_{k}N)+N^{\dagger}\sigma_{i}N\nabla_{j}(N^{\dagger}\tau_{3}\sigma_{k}N)]\\\nonumber
&-\tilde{{\mathcal{G}}}_{5}{\mathcal{I}}_{ab}\epsilon_{ijk}N^{\dagger}\tau_{a}\sigma_{i}N\nabla_{j}(N^{\dagger}\tau_{b}\sigma_{k}N)+{\mathcal{G}}_{6}\epsilon_{ab3}\stackrel{\rightarrow}{\nabla}(N^{\dagger}\tau_{a}N)\cdot N^{\dagger}\tau_{b}\vectS{\sigma}N
\end{align}

\noindent (In Eq. (\ref{eq:GirlandaLagrangian}) a factor of $1/\Lambda_{\chi}^{3}$ has been absorbed into the coefficients.  This notation agrees with the notation of Phillips, Schindler, and Springer\cite{Phillips:2008hn}.)  With this parity-violating Lagrangian one can compute the Girlanda potential which takes on the following form given by Eq. (\ref{eq:PVpotential}), where $n$ runs from one to five.

\begin{table}[h]
\caption{\label{tbl:Girlandapvpotential}Parity violating potential in Girlanda formalism.  ${\cal T}_{ij}\equiv (3\tau_i^z\tau_j^z-\tau_i\cdot\tau_j)$.
}
\begin{ruledtabular}
\begin{center}
\begin{tabular}{cccc}
$n$ & $c_{n}^{Gir}$ & $f_{n}^{Gir}(r)$ &$O^{(n)}_{ij}$ \\
\hline
$1$ & $2{\mathcal{G}}_{6}$ & $\delta^{3}(\vect{r})$ &
      $(\tau_i\times\tau_j)^z(\vectS{\sigma}_i+\vectS{\sigma}_j)\cdot{\boldsymbol{X}}^{(1)}_{ij,-}$
\\
$2 $ &  ${\mathcal{G}}_{2}$ & $\delta^{3}(\vect{r})$ &
      $(\tau_i+\tau_j)^z(\vectS{\sigma}_i-\vectS{\sigma}_j)\cdot{\boldsymbol{X}}^{(2)}_{ij,+}$
\\
$3 $ & $-2{\mathcal{G}}_{5}$ & $\delta^{3}(\vect{r})$ &
  ${\cal T}_{ij}
   (\vectS{\sigma}_i-\vectS{\sigma}_j)\cdot{\boldsymbol{X}}^{(3)}_{ij,+}$
\\
$4 $ & $2{\mathcal{G}}_{1}$ & $\delta^{3}(\vect{r})$ &
     $(\vectS{\sigma}_i-\vectS{\sigma}_j)\cdot{\boldsymbol{X}}^{(4)}_{ij,+}$
\\
$5  $ & $2\tilde{\mathcal{G}}_{1}$ & $\delta^{3}(\vect{r})$ &
    $  (\vectS{\sigma}_i\times\vectS{\sigma}_j)\cdot{\boldsymbol{X}}^{(5)}_{ij,-}$
\end{tabular}
\end{center}
\end{ruledtabular}
\end{table} 

\noindent Using (\ref{eq:GirlandaResults}) one can reduce the Zhu potential to a set of five operators, allowing the matching of the Zhu coefficients onto the Girlanda coefficients as shown in Table \ref{tbl:dictionary}.

For our calculations, we also require the coefficients in the auxiliary field formalism Eq. (\ref{eq:AuxLagr}).  This matching of the $\mathcal{G}_{i}$ and $g_{i}$ requires two steps.  One first performs Gaussian integration over the auxiliary fields followed by a field redefinition to rewrite the Lagrangian, Eq (\ref{eq:AuxLagr}) in terms of nucleon fields, as done by Schindler, and Springer\cite{Schindler:2009wd}.  Then one can match this partial wave formalism onto the Girlanda formalism by performing Fierz rearrangements and using the constraints  Eq. (\ref{eq:GirlandaResults}), yielding the results in Table \ref{tbl:dictionary}  (This has also been done using a different method by Phillips, Schindler, and Springer\cite{Phillips:2008hn}.)  

Finally we wish to match the Girlanda potential onto the Danilov potential which is given by Eq. (\ref{eq:PVpotential}) in Table \ref{tbl:Danilovpvpotential}, where $n$ runs from one to five.

\begin{table}[h]
\caption{\label{tbl:Danilovpvpotential}Parity violating potential in Danilov formalism.  ${\cal T}_{ij}\equiv (3\tau_i^z\tau_j^z-\tau_i\cdot\tau_j)$, $P_{0}=\frac{1}{4}\left(1-\vectS{\sigma}_{i}\cdot\vectS{\sigma}_{j}\right)$, $P_{1}=\frac{1}{4}\left(3+\vectS{\sigma}_{i}\cdot\vectS{\sigma}_{j}\right)$.
}
\begin{ruledtabular}
\begin{center}
\begin{tabular}{cccc}
$n$ & $c_{n}^{Dan}$ & $f_{n}^{Dan}(r)$ &$O^{(n)}_{ij}$ \\
\hline
$1$ & $\frac{1}{2}\rho_{t}/\gamma_{d}$ & $\frac{4\pi}{M_{N}}\delta^{3}(\vect{r})$ &
      $(\tau_i-\tau_j)^z(\vectS{\sigma}_i+\vectS{\sigma}_j)\cdot{\boldsymbol{X}}^{(1)}_{ij,-}$
\\
$2 $ &  $\frac{1}{2}\lambda_{s}^{1}/\gamma_{t}$ & $\frac{4\pi}{M_{N}}\delta^{3}(\vect{r})$ &
      $(\tau_i+\tau_j)^z(\vectS{\sigma}_i-\vectS{\sigma}_j)\cdot{\boldsymbol{X}}^{(2)}_{ij,+}$
\\
$3 $ & $\frac{1}{2\sqrt{6}}\lambda_{s}^{2}/\gamma_{t}$ & $\frac{4\pi}{M_{N}}\delta^{3}(\vect{r})$ &
  ${\cal T}_{ij}
   (\vectS{\sigma}_i-\vectS{\sigma}_j)\cdot{\boldsymbol{X}}^{(3)}_{ij,+}$
\\
$4 $ & $\lambda_{t}/\gamma_{d}$ & $\frac{4\pi}{M_{N}}\delta^{3}(\vect{r})$ &
     $(\vectS{\sigma}_i-\vectS{\sigma}_j)P_{1}\cdot{\boldsymbol{X}}^{(4)}_{ij,+}$
\\
$5  $ & $\lambda_{s}^{0}/\gamma_{t}$ & $\frac{4\pi}{M_{N}}\delta^{3}(\vect{r})$ &
    $  (\vectS{\sigma}_i-\vectS{\sigma}_j)P_{0}\cdot{\boldsymbol{X}}^{(5)}_{ij,-}$
\end{tabular}
\end{center}
\end{ruledtabular}
\end{table} 

\noindent In order to match the Girlanda formalism to the Danilov formalism we note the identity

\begin{equation}
\label{eq:StoP}
\langle P|[-i\boldsymbol{\nabla},\delta^{3}(\vect{r})]|S\rangle=\langle P|\left\{-i\boldsymbol{\nabla},\delta^{3}(\vect{r})\right\}|S\rangle
\end{equation}

\noindent which follows trivially since P waves are zero at the origin.  Next we make use of the identical identities in spin and isospin space.  (Note $P_{0}^{\tau}=\frac{1}{4}\left(1-\vectS{\tau_{i}}\cdot\vectS{\tau_{j}}\right)$ and $P_{1}^{\tau}=\frac{1}{4}\left(3+\vectS{\tau_{i}}\cdot\vectS{\tau_{j}}\right)$)

\begin{equation}
\label{eq:Spin}
i(\vectS{\sigma}_{i}\times\vectS{\sigma}_{j})=(\vectS{\sigma}_{i}-\vectS{\sigma}_{j})\left(P_{0}-P_{1}\right)
\end{equation}

\begin{equation}
\label{eq:Iso}
i(\vectS{\tau}_{i}\times\vectS{\tau}_{j})^{z}=(\vectS{\tau}_{i}-\vectS{\tau}_{j})^{z}\left(P_{0}^{\tau}-P_{1}^{\tau}\right)
\end{equation}

\noindent The isospin operator $i(\vectS{\tau}_{i}\times\vectS{\tau}_{j})^{z}$ only appears with the spin operator $(\vectS{\sigma}_{i}+\vectS{\sigma}_{j})$ in the Girlanda potential.  Since this spin operator only projects out the triplet state of the S wave, only the isosinglet part of the operator $i(\vectS{\tau}_{i}\times\vectS{\tau}_{j})^{z}$ is projected out.  Thus by Eq. (\ref{eq:Iso}) we find that in combination with the spin operator $(\vectS{\sigma}_{i}+\vectS{\sigma}_{j})$ the isospin operator $i(\vectS{\tau}_{i}\times\vectS{\tau}_{j})^{z}=(\vectS{\tau}_{i}-\vectS{\tau}_{j})^{z}$.  Finally using Eqs. (\ref{eq:StoP}), (\ref{eq:Spin}), and the fact that the identity $I$ is $I=P_{0}+P_{1}$ one can straightforwardly match the Girlanda coefficients to the Danilov coefficients, giving the results shown in Table \ref{tbl:dictionary}.  Also shown in Table \ref{tbl:dictionary} are the relation between the Zhu, Girlanda, Auxiliary, and Danilov formalisms.  The primary goal in low energy hadronic parity violation is to determine the value of the Danilov parameters.  At low energies all of these different EFT formalisms can be shown to be equivalent to the Danilov parameters, as shown in Table \ref{tbl:dictionary}.  Thus one can use whichever formalism is more convenient. 

\begin{table}[h]
\caption{\label{tbl:dictionary}Translation between various formalisms of parity violating potential
}
\begin{center}
\begin{tabular}{|l|l|l|l|}
\hline
\multicolumn{1}{|c|}{Zhu} & \multicolumn{1}{|c|}{Girlanda} & Auxiliary & Danilov  \\
\hline
$\frac{M_{N}\gamma_{d}}{2\pi\Lambda_{\chi}^{3}}\left(C_{1}-\tilde{C}_{1}-3(C_{3}-\tilde{C}_{3})\right)$
&
$\frac{M_{N}\gamma_{d}}{2\pi}\left(\tilde{\mathcal{G}}_{1}-{\mathcal{G}}_{1}\right) $ & $-2\sqrt{2}g_{1}$ & $\lambda_{t}$ \\
$\frac{M_{N}\gamma_{d}}{\pi\Lambda_{\chi}^{3}}\left(\tilde{C}_{6}+\frac{1}{2}(C_{2}-C_{4})\right)$ & $\frac{M_{N}\gamma_{d}}{\pi}{\mathcal{G}}_{6}$ & $-4\sqrt{2}g_{2}$ & $\rho_{t}$ \\
$\frac{M_{N}\gamma_{t}}{2\pi\Lambda_{\chi}^{3}}\left(C_{1}+\tilde{C}_{1}+(C_{3}+\tilde{C}_{3})\right) $ &
$\frac{M_{N}\gamma_{t}}{2\pi}\left(\tilde{\mathcal{G}}_{1}+{\mathcal{G}}_{1}\right)
$ & $-2\sqrt{2}g_{3}$ & $\lambda_{s}^{0}$ \\
$\frac{M_{N}\gamma_{t}}{2\pi\Lambda_{\chi}^{3}}\left(C_{2}+C_{4}+\tilde{C}_{2}+\tilde{C}_{4}\right) $ & $\frac{M_{N}\gamma_{t}}{2\pi}{\mathcal{G}}_{2}$ & $-\sqrt{2}g_{4}$ & $\lambda_{s}^{1}$ \\
$-\frac{M_{N}\gamma_{t}\sqrt{6}}{\pi\Lambda_{\chi}^{3}}\left(C_{5}+\tilde{C}_{5}\right)$ & $-\frac{M_{N}\gamma_{t}\sqrt{6}}{\pi}{\mathcal{G}}_{5} $ & $-4\sqrt{3}g_{5}$
& $\lambda_{s}^{2}$  \\
\hline
\end{tabular}
\end{center}
\end{table} 

Having matched the auxiliary coefficients $g_{i}$ to the Zhu coefficients we can now use the matching of the Zhu coefficients to the DDH ``best" values to obtain estimates for the auxiliary coefficients which yields.

\begin{subequations}
\begin{align}
g_{1}&=-\frac{M_{N}\gamma_{d}}{8\sqrt{2}\pi}\left[\frac{g_{\omega}\chi_{\omega}}{M_{N}m_{\omega}^{2}}h_{\omega}^{0}-\frac{3g_{\rho}\chi_{\rho}}{M_{N}m_{\rho}^{2}}h_{\rho}^{0}\right]\sim-7.89\times10^{-11}\mathrm{MeV}^{-1}\\
g_{2}&=\frac{M_{N}\gamma_{d}}{4\sqrt{2}\pi}\left[-\frac{g_{\pi NN}}{4\sqrt{2}M_{N}m_{\pi}^{2}}f_{\pi}-\frac{g_{\rho}}{4M_{N}m_{\rho}^{2}}h_{\rho}^{1}+\frac{g_{\omega}}{4M_{N}m_{\omega}^{2}}h_{\omega}^{1}\right]\sim-1.43\times10^{-10}\mathrm{MeV}^{-1}\\
g_{3}&=\frac{M_{N}\gamma_{t}}{8\sqrt{2}\pi}\left[\frac{g_{\omega}(2+\chi_{\omega})}{M_{N}m_{\omega}^{2}}h_{\omega}^{0}+\frac{g_{\rho}(2+\chi_{\rho})}{M_{N}m_{\rho}^{2}}h_{\rho}^{0}\right]\sim8.53\times10^{-12}\mathrm{MeV}^{-1}\\
g_{4}&=\frac{M_{N}\gamma_{t}}{4\sqrt{2}\pi}\left[\frac{g_{\rho}(2+\chi_{\rho})}{M_{N}m_{\rho}^{2}}h_{\rho}^{1}+\frac{g_{\omega}(2+\chi_{\omega})}{M_{N}m_{\omega}^{2}}h_{\omega}^{1}\right]\sim1.67\times10^{-12}\mathrm{MeV}^{-1}\\
g_{5}&=\frac{M_{N}\gamma_{t}}{8\sqrt{2}\pi}\left[\frac{g_{\rho}(2+\chi_{\rho})}{\sqrt{6}M_{N}m_{\rho}^{2}}h_{\rho}^{2}\right]\sim2.50\times10^{-12}\mathrm{MeV}^{-1}
\end{align}
\end{subequations}


\section{Spin Observables}

Having calculated the various parity-violating amplitudes, we can now relate them to PV observables.  One such observable is the neutron longitudinal asymmetry $A_{N}$\cite{Schiavilla:2004wn}.  In this case we scatter longitudinally polarized neutrons from an unpolarized deuteron target, and measure the difference of the two cross sections.  

\begin{equation}
A_{N}=\frac{\sigma_{+}-\sigma_{-}}{\sigma_{+}+\sigma_{-}}
\end{equation}

\noindent Here $\sigma_{+}$ ($\sigma_{-}$) represents the cross section of positive (negative) helicity neutrons.  \newline\indent In order to calculate observables, we need to write them in terms of the partial wave amplitudes calculated above. We denote the transition matrix by the operator $\mathbf{M}$. Of course, $\mathbf{M}$ is not diagonal in the orbital angular momentum basis, but rather is diagonal in terms of total angular momentum.  Defining $M_{m_{1}',m_{2}';m_{1},m_{2}}$ as the $T$ matrix where the nucleon has initial (final) spin $m_{2}$, ($m_{2}'$), and the deuteron has initial (final) spin $m_{1}$, ($m_{1}'$),it can be shown

\begin{align}
\label{eq:TmatrixProjection} M_{m_{1}',m_{2}';m_{1},m_{2}}=&\sqrt{4\pi}\sum_{J}\sum_{L,L'}\sum_{S,S'}\sum_{m_{s},m_{S}'}\sum_{m_{L}'}\sqrt{2L+1}C_{1,\nicefrac{1}{2},S}^{m_{1},m_{2},m_{S}}\\\nonumber
&C_{1,\nicefrac{1}{2},S'}^{m_{1}',m_{2}',m_{S}'}C_{L,S,J}^{0,m_{S},M}C_{L',S',J}^{m_{L}',m_{S}',M}Y_{L'}^{m_{L}'}(\theta,\phi)M^{J}_{L'S',LS}
\end{align}

\noindent Observables are most easily written in terms of this matrix $M_{m_{1}',m_{2}';m_{1},m_{2}}$.  Having Eq. (\ref{eq:TmatrixProjection}), which gives $M_{m_{1}',m_{2}';m_{1},m_{2}}$  in terms of the calculated functions $M^{J}_{L'S',LS}$, we can calculate observables by truncating the sum over $J,L,$ and $L'$ at some reasonable level.  Thus the observable $A_{N}$ is given by



\begin{align}
A_{N}=\frac{\sum\limits_{m_{1}',m_{2}'}\sum\limits_{m_{1},m_{2}}(-1)^{\nicefrac{1}{2}-m_{2}}\int d\Omega|M_{m_{1}',m_{2}';,m_{2},m_{2}}|^{2}}{\sum\limits_{m_{1}',m_{2}'}\sum\limits_{m_{1},m_{2}}\int d\Omega|M_{m_{1}',m_{2}';,m_{2},m_{2}}|^{2}}
\end{align}

\noindent Explicitly summing over values of angular momentum from 0 to 1, and spin and $J$ from $J=\nicefrac{1}{2}$ to $J=\nicefrac{3}{2}$ we find.

\begin{align}
A_{N}\cong&\frac{2}{3}\mathrm{Re}\left[\left(M^{\nicefrac{1}{2}}_{0\nicefrac{1}{2},0\nicefrac{1}{2}}+M^{\nicefrac{1}{2}}_{1\nicefrac{1}{2},1\nicefrac{1}{2}}\right)\left(M^{\nicefrac{1}{2}}_{1\nicefrac{1}{2},0\nicefrac{1}{2}}\right)^{*}+2\sqrt{2}\left(M^{\nicefrac{1}{2}}_{0\nicefrac{1}{2},0\nicefrac{1}{2}}+M^{\nicefrac{1}{2}}_{1\nicefrac{3}{2},1\nicefrac{3}{2}}\right)\left(M^{\nicefrac{1}{2}}_{1\nicefrac{3}{2},0\nicefrac{1}{2}}\right)^{*}\right.\\\nonumber
&\left.-4\left(M^{\nicefrac{3}{2}}_{0\nicefrac{3}{2},0\nicefrac{3}{2}}+M^{\nicefrac{1}{2}}_{1\nicefrac{1}{2},1\nicefrac{1}{2}}\right)\left(M^{\nicefrac{3}{2}}_{1\nicefrac{1}{2},0\nicefrac{3}{2}}\right)^{*}-2\sqrt{5}\left(M^{\nicefrac{3}{2}}_{0\nicefrac{3}{2},0\nicefrac{3}{2}}+M^{\nicefrac{3}{2}}_{1\nicefrac{3}{2},1\nicefrac{3}{2}}\right)\left(M^{\nicefrac{3}{2}}_{1\nicefrac{3}{2},0\nicefrac{3}{2}}\right)^{*}\right]/\\\nonumber
&\left[|M^{\nicefrac{1}{2}}_{0\nicefrac{1}{2},0\nicefrac{1}{2}}|^{2}+2|M^{\nicefrac{3}{2}}_{0\nicefrac{3}{2},0\nicefrac{3}{2}}|^{2}+3|M^{\nicefrac{1}{2}}_{1\nicefrac{1}{2},1\nicefrac{1}{2}}|^{2}+3|M^{\nicefrac{3}{2}}_{1\nicefrac{3}{2},1\nicefrac{3}{2}}|^{2}\right]
\end{align}

\noindent By the optical theorem we note that $A_{N}$ can also be written as 

\begin{align}
A_{N}&=\frac{\sum_{m}\mathrm{Im}\left(M_{m,\nicefrac{1}{2};m,\nicefrac{1}{2}}|_{\theta=0}-M_{m,-\nicefrac{1}{2};m,-\nicefrac{1}{2}}|_{\theta=0}\right)}{\sum_{m}\mathrm{Im}\left(M_{m,\nicefrac{1}{2};m,\nicefrac{1}{2}}|_{\theta=0}+M_{m,-\nicefrac{1}{2};m,-\nicefrac{1}{2}}|_{\theta=0}\right)}\\\nonumber
&\cong\frac{2}{3}\mathrm{Im}\left[M^{\nicefrac{1}{2}}_{1\nicefrac{1}{2},0\nicefrac{1}{2}}+2\sqrt{2}M^{\nicefrac{1}{2}}_{1\nicefrac{3}{2},0\nicefrac{1}{2}}-4M^{\nicefrac{3}{2}}_{1\nicefrac{1}{2},0\nicefrac{3}{2}}-2\sqrt{5}M^{\nicefrac{3}{2}}_{1\nicefrac{3}{2},0\nicefrac{1}{2}}\right]/\\\nonumber
&\quad\quad\,\mathrm{Im}\left[M^{\nicefrac{1}{2}}_{0\nicefrac{1}{2},0\nicefrac{1}{2}}+2M^{\nicefrac{3}{2}}_{0\nicefrac{3}{2},0\nicefrac{3}{2}}+3M^{\nicefrac{1}{2}}_{1\nicefrac{1}{2},1\nicefrac{1}{2}}+3M^{\nicefrac{1}{2}}_{1\nicefrac{3}{2},1\nicefrac{3}{2}}\right]
\end{align}

Another parity-violating observable is the spin rotation of the neutron as it passes through a deuteron target.  In this experiment the neutron is transversely polarized and the rate of change of the rotation angle with respect to the distance traveled is\cite{Schiavilla:2008ic,PhysRevC.83.029902,Schiavilla:2004wn}.

\begin{equation}
\frac{d\phi}{dz}=-\frac{4M_{N} N}{9k}\sum_{m}\mathrm{Re}\left[M_{m,\nicefrac{1}{2};m,\nicefrac{1}{2}}|_{\theta=0}-M_{m,-\nicefrac{1}{2};m,-\nicefrac{1}{2}}|_{\theta=0}\right]
\end{equation}

\noindent where $N$ is the number of scattering centers per units volume, and $k$ is the momentum of the neutron in the c.m. system.  Using Eq. (\ref{eq:TmatrixProjection}) we find

\begin{equation}
\label{eq:SpinRotation}
\frac{d\phi}{dz}=-\frac{4M_{N}N}{27k}\mathrm{Re}\left[M^{\nicefrac{1}{2}}_{1\nicefrac{1}{2},0\nicefrac{1}{2}}+2\sqrt{2}M^{\nicefrac{1}{2}}_{1\nicefrac{3}{2},0\nicefrac{1}{2}}-4M^{\nicefrac{3}{2}}_{1\nicefrac{1}{2},0\nicefrac{3}{2}}-2\sqrt{5}M^{\nicefrac{3}{2}}_{1\nicefrac{3}{2},0\nicefrac{3}{2}}\right]
\end{equation}

The final parity-violating observable that we will consider is the deuteron target asymmetry.  In this case an unpolarized beam of neutrons is scattered from a polarized deuteron target.  At first the deuteron target is polarized in the positive $z$ direction, where  $\hat{z}$ is the direction of the neutron's initial momentum.  Then the neutron scatters off the deuteron target and we measure the cross section $\sigma_{1}$.  In addition the deuteron target is polarized in the opposite direction and we measure the cross section $\sigma_{-1}$.  If there is parity violation one will find that $\sigma_{1}\neq\sigma_{-1}$, and we define the target asymmetry as  

\begin{equation}
A_{D}=\frac{\sigma_{1}-\sigma_{-1}}{\sigma_{1}+\sigma_{-1}}
\end{equation}

\begin{align}
A_{D}=&\frac{\sum\limits_{m_{1'},m_{2}'}\sum\limits_{m_{2}}\int d\Omega\left(|M_{m_{1}',m_{2}';1,m_{2}}|^{2}-|M_{m_{1}',m_{2}';-1,m_{2}}|^{2}\right)}{\sum\limits_{m_{1'},m_{2}'}\sum\limits_{m_{2}}\int d\Omega\left(|M_{m_{1}',m_{2}';1,m_{2}}|^{2}+|M_{m_{1}',m_{2}';-1,m_{2}}|^{2}\right)}\\\nonumber
\cong&-\mathrm{Re}\left[2\left(M_{1\nicefrac{1}{2},1\nicefrac{1}{2}}^{\nicefrac{1}{2}}+M_{0\nicefrac{1}{2},0\nicefrac{1}{2}}^{\nicefrac{1}{2}}\right)\left(M_{1\nicefrac{1}{2},0\nicefrac{1}{2}}^{\nicefrac{1}{2}}\right)^{*}+\sqrt{2}\left(M_{1\nicefrac{3}{2},1\nicefrac{3}{2}}^{\nicefrac{1}{2}}+M_{0\nicefrac{1}{2},0\nicefrac{1}{2}}^{\nicefrac{1}{2}}\right)\left(M_{1\nicefrac{3}{2},0\nicefrac{1}{2}}^{\nicefrac{1}{2}}\right)^{*}\right.\\\nonumber
&\left.-2\left(M_{1\nicefrac{1}{2},1\nicefrac{1}{2}}^{\nicefrac{3}{2}}+M_{0\nicefrac{3}{2},0\nicefrac{3}{2}}^{\nicefrac{3}{2}}\right)\left(M_{1\nicefrac{1}{2},0\nicefrac{3}{2}}^{\nicefrac{3}{2}}\right)^{*}+2\sqrt{5}\left(M_{1\nicefrac{3}{2},1\nicefrac{3}{2}}^{\nicefrac{3}{2}}+M_{0\nicefrac{3}{2},0\nicefrac{3}{2}}^{\nicefrac{3}{2}}\right)\left(M_{1\nicefrac{3}{2},0\nicefrac{3}{2}}^{\nicefrac{3}{2}}\right)^{*}\right]/\\\nonumber
&\left[|M_{0\nicefrac{1}{2},0\nicefrac{1}{2}}^{\nicefrac{1}{2}}|^{2}+2|M_{0\nicefrac{3}{2},0\nicefrac{3}{2}}|^{2}+3|M_{1\nicefrac{1}{2},1\nicefrac{1}{2}}^{\nicefrac{1}{2}}|^{2}+6|M_{1\nicefrac{3}{2},1\nicefrac{3}{2}}^{\nicefrac{1}{2}}|^{2}\right]
\end{align}

\noindent Again $A_{D}$ can also be calculated via the optical theorem.

\begin{align}
A_{D}=&\frac{\sum_{m_{2}}\mathrm{Im}\left(M_{1,m_{2};1,m_{2}}|_{\theta=0}-M_{-1,m_{2};-1,m_{2}}|_{\theta=0}\right)}{\sum_{m_{2}}\mathrm{Im}\left(M_{1,m_{2};1,m_{2}}|_{\theta=0}+M_{-1,m_{2};-1,m_{2}}|_{\theta=0}\right)}\\\nonumber
\cong&-\mathrm{Im}\left[2M_{1\nicefrac{1}{2},0\nicefrac{1}{2}}^{\nicefrac{1}{2}}+\sqrt{2}M_{1\nicefrac{3}{2},0\nicefrac{1}{2}}^{\nicefrac{1}{2}}-2M_{1,\nicefrac{1}{2},0\nicefrac{3}{2}}^{\nicefrac{3}{2}}+2\sqrt{5}M_{1\nicefrac{3}{2},0\nicefrac{3}{2}}^{\nicefrac{3}{2}}\right]/\\\nonumber
&\mathrm{Im}\left[M_{0\nicefrac{1}{2},0\nicefrac{1}{2}}^{\nicefrac{1}{2}}+2M_{0\nicefrac{3}{2},0\nicefrac{3}{2}}^{\nicefrac{3}{2}}+3M_{1\nicefrac{1}{2},1\nicefrac{1}{2}}^{\nicefrac{1}{2}}+6M_{1\nicefrac{3}{2},1\nicefrac{3}{2}}^{\nicefrac{1}{2}}\right]
\end{align}


\section{Results}

\begin{figure}[h]
 \centering
 	\begin{tabular}{cc}
	 	\newline
 		\includegraphics[width=90mm]{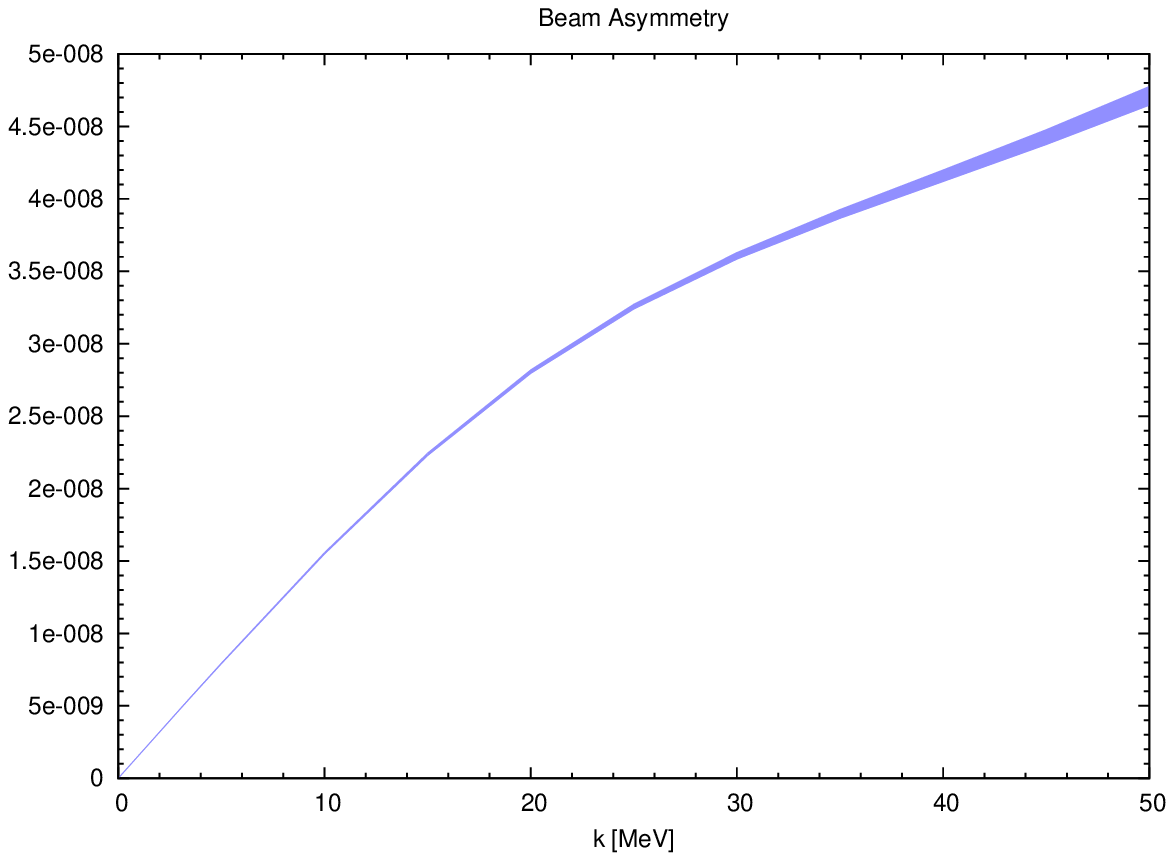} & \includegraphics[width=90mm]{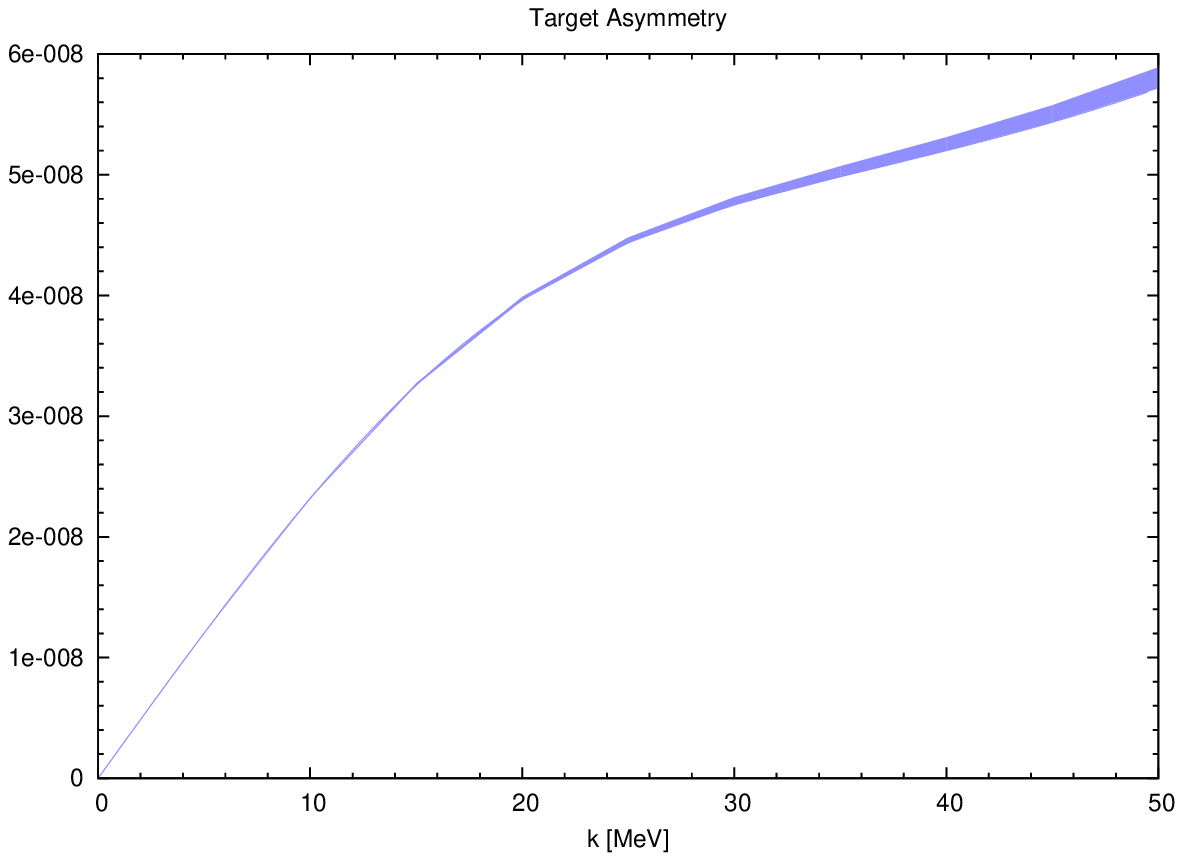}\\
 	\end{tabular}	
\caption{\label{fig:observables}Beam and target asymmetries as function of c.m. momentum k}
\end{figure}

Plotting our results for beam and target asymmetry as a function of center of mass momentum $k$ we find the plots given in Fig. \ref{fig:observables}.  The thickness of the plot denotes the cutoff variation.  The cutoff variation runs from 200 MeV to infinity.  It appears that the results begin to converge after 900 MeV as found by other authors\cite{Griesshammer:2004pe}.  Also it should be noted that the cutoff variation for the beam and target asymmetries at low energies is actually smaller than as shown in the plots, and is displayed with the given thickness in order that the plot be visible.  The cutoff variation of the spin rotation which is not plotted, is more significant than for $A_{N}$ and $A_{D}$, which should come as no surprise since we are dividing by a factor of momentum in Eq. (\ref{eq:SpinRotation}).  (We use a liquid deuterium density of $N=.4\times 10^{23} \mathrm{atoms}\ \mathrm{cm}^{-3}$\cite{Song:2010sz}.)  This means that the spin rotation is sensitive to higher order effects and, since we only did a LO calculation, the cutoff sensitivity should be much greater.  The plots for the beam and target asymmetries extend all the way to 50 MeV.  However, the observables cannot be taken seriously at these high energies as higher partial waves and higher order contributions will become important.  (It should also be noted that a significant difference was found at about 20 MeV if P waves were not included in the parity-conserving amplitudes.)  The spin rotation observable at zero energy ranged from $6.57\times 10^{-9}$ $\mathrm{rad}\ \mathrm{cm}^{-1}$ to $6.82\times 10^{-9}$ $\mathrm{rad}\ \mathrm{cm}^{-1}$ due to cutoff dependence.  This order of magnitude falls in line with previous calculations\cite{Schiavilla:2008ic,PhysRevC.83.029902,Song:2010sz}.  Also the beam asymmetry ranges from $6.79\times 10^{-9}$ to $6.83\times 10^{-9}$ due to cutoff dependence, at $E_{lab}=15$ keV, and this also agrees well with previous calculations\cite{Song:2010sz}.  Finally the target asymmetry varies little due to cutoff dependence at $E_{lab}=15$ keV with a value of roughly $7.84\times 10^{-9}$.

Finally in order to compare with possible experiments a table of all three observables in terms of their contributions from each of the $g_{i}$ is given below.  The spin rotation is given at zero energy and the beam and target asymmetry are given at a lab energy of 15 keV.  In order to obtain a prediction for the observable each row is multiplied by the appropriate value of $g_{i}$ and then these products are added together to yield the observable.  Below are two tables with different values for the cutoff.  The first table shows the cutoff at 200 MeV and the second at setting the cutoff to infinity.

\begin{table}[h]
\caption{\label{tbl:values_200} value for cutoff $\Lambda=200$ MeV
}
\begin{center}
\begin{tabular}{|c|c|c|c|}
\hline
$g_{i}$ & Rotation, $E_{lab}=0$ keV & $A_{N}$, $E_{lab}=15$ keV & $A_{D}$, $E_{lab}=15$ keV\\\hline
1 &-9.33 rad $\mathrm{cm}^{-1}$ MeV &-14.4 MeV &8.92 MeV \\
2 & -18.1 rad $\mathrm{cm}^{-1}$ MeV& -39.6 MeV&-59.7 MeV \\
3 & -5.11 rad $\mathrm{cm}^{-1}$ MeV&-1.83 MeV& 1.65 MeV\\
4 & 3.40 rad $\mathrm{cm}^{-1}$ MeV&1.22 MeV& -1.10 MeV\\
\hline
\end{tabular}
\end{center}
\end{table}

\begin{table}[h]
\caption{\label{tbl:values_1000}value for cutoff $\Lambda=\infty$ MeV
}
\begin{center}
\begin{tabular}{|c|c|c|c|}
\hline
$g_{i}$ & Rotation, $E_{lab}=0$ keV & $A_{N}$, $E_{lab}=15$ keV & $A_{D}$, $E_{lab}=15$ keV\\\hline
1 &-9.58 rad $\mathrm{cm}^{-1}$ MeV &-14.5 MeV &9.15 MeV \\
2 & -19.0 rad $\mathrm{cm}^{-1}$ MeV& -39.9 MeV&-59.8 MeV \\
3 & -8.35 rad $\mathrm{cm}^{-1}$ MeV&-2.71 MeV & 2.47 MeV\\
4 & 5.56 rad $\mathrm{cm}^{-1}$ MeV&1.81 MeV & -1.65 MeV\\
\hline
\end{tabular}
\end{center}
\end{table}

 Finally we should note that the calculation for the beam and target asymmetries were done using both the standard cross section methods and the optical theorem.  Plotting the results from both, we found they were indistinguishable.  This agreement confirms that our amplitudes are unitary and acts as a check on the validity of our results.  For the values quoted in the table, it was found that for the beam and target asymmetries, as well as the values from either the cross sections or optical theorem agreed to less than one percent.

\section{Conclusion}
Above we calculated the low energy parity-violating $nd$ transition amplitudes using pionless EFT.  Matching the auxiliary field formalism onto the DDH potential, we made predictions for the coefficients of the auxiliary field formalism.  Using these amplitudes, we were able to make predictions for the spin rotation, beam asymmetry, and target asymmetry in low energy $nd$ interactions.  The values obtained for the neutron spin rotation and beam asymmetry are in agreement with values found by other authors\cite{Schiavilla:2008ic,PhysRevC.83.029902,Song:2010sz}.  Unfortunately due to the smallness of these values they will require very precise experiments to measure.  However, the five LEC's used in parity-violation at LO are not very well determined.  Thus upon further experiments it may be found that the values of the LEC's are such that the observables for $nd$ interactions are larger than predicted.

The largest contribution to parity violation was shown to come from the coefficient $g_{2}$, which contains the pion contribution, and such experiments should then allow one to determine its value.  It is noted that to first order in parity violation the $\Delta I=2$ ($g_{5}$) term does not contribute.  Thus $nd$ scattering is sensitive to four out of the five PV coefficients.

In principle we should be able to calculate to NLO in \EFT without the need for parity-violating three-body forces\cite{Griesshammer:2010nd}.  Griesshammer, Schindler, and Springer calculated the NLO parity violating amplitudes using the partially resummed approach which introduces higher order contributions at NLO\cite{Griesshammer:2011md}.  However, to calculate the NLO contributions without higher order contributions, one must calculate the full off shell LO amplitude.  Since a calculation of the full off shell LO amplitude is numerically expensive it will be left to a future publication.

\appendix
\numberwithin{equation}{section}

\section{Appendix}

In projecting out the amplitudes one has to perform angular integrations which are given by

\begin{align}
\label{eq:AngularProjection2}
U_{JL}=&\sqrt{\frac{4\pi}{3}}\int\!d\Omega_{k}\int\!d\Omega_{p}\frac{1}{a+\hat{\mathbf{k}}\cdot\hat{\mathbf{p}}}{Y_{L'}^{m_{L}'}}^{*}(\hat{\mathbf{p}})Y_{L}^{m_{L}}(\hat{\mathbf{k}})\left(kY_{1}^{m}(\hat{\mathbf{k}})+2pY_{1}^{m}(\hat{\mathbf{p}})\right)=\\\nonumber
&=4\pi\sqrt{\frac{2L+1}{2L'+1}}C_{L,1,L'}^{0,0,0}C_{L,1,L'}^{m_{L},m,m_{L}'}(kQ_{L'}(a)+2pQ_{L}(a))
\end{align}

\noindent where 

\begin{equation}
Q_{L}=\frac{1}{2}\int_{-1}^{1}\frac{P_{L}(x)}{x+a}dx
\end{equation}

\noindent are functions related to the Legendre polynomials of the second kind up to a factor of $(-1)^{L}$, and $P_{L}(x)$ are the standard Legendre polynomials.

The projection we must carry out is of the form 

\begin{equation}
\label{eq:KProj}
\boldsymbol{\mathcal{K}}(k,p)^{J}_{L'S',LS}=\int d\Omega_{k}\int d\Omega_{p} \left({\mathcal{Y}}^{M}_{J,L'S'}(\hat{\mathbf{p}})\right)^{*}\left(\boldsymbol{\mathcal{K}}^{ji}\right)^{\beta b}_{\alpha a}(\vect{k},\vect{p}){\mathcal{Y}}^{M}_{J,LS}(\hat{\mathbf{k}})
\end{equation}

\noindent (Note the polarization and spin indices are summed over corresponding indices that are not explicitly shown in the spin angle functions). Each matrix element of $\left(\boldsymbol{\mathcal{K}}^{ji}\right)^{\beta b}_{\alpha a}(\vect{k},\vect{p})$ has a different projection.  Each of these four different projections has in turn four pieces given by Eqs. (\ref{seq:Kequations}).  Fortunately many of these terms can be related by time reversal simplifying the projection considerably.  For simplicity we only show how to project out the first $g^{{}^{3}\!S_{1}-{}^{3}\!P_{1}}$ piece of the matrix element $\left[\left(\boldsymbol{\mathcal{K}}^{ji}\right)^{\beta b}_{\alpha a}(\vect{k},\vect{p})\right]_{11}$  as given in (\ref{eq:Krelation}) and (\ref{eq:K11}), and simply quote the other results.  In order to project this term out we use properties of spherical tensors and the Wigner-Eckart theorem to reduce the following expression to a sum over Clebsch-Gordan coefficients.

\begin{align}
W_{JL}=&i\epsilon^{i\ell k}\langle\nicefrac{1}{2},m_{2}'|\sigma^{k}\sigma^{j}|\nicefrac{1}{2},m_{2}\rangle(\vect{k}+2\vect{p})^{\ell}\\\nonumber
=&\sqrt{\frac{2}{3}}\sum_{m,m'}\sum_{k,q}\sqrt{2k+1}C_{1,1,1}^{m_{1},m,m'}C_{1,k,1}^{m_{1}',q,m'}C_{\nicefrac{1}{2},k,\nicefrac{1}{2}}^{m_{2},q,m_{2}'}\langle\nicefrac{1}{2}||T_{k}||\nicefrac{1}{2}\rangle(-1)^{m}(\vect{k}+2\vect{p})^{-m}
\end{align} 

\noindent Using the above expression with (\ref{eq:Krelation}),(\ref{eq:K11}), and (\ref{eq:KProj}) and, for the time being ignoring the isospin, we find the expression for our projection is.

\begin{align}
V_{JL}=&\sqrt{\frac{4\pi}{3}}\frac{1}{kp}g^{{}^{3}\!S_{1}-{}^{3}\!P_{1}}\int\! d\Omega_{p}\int\! d\Omega_{k}\frac{1}{a+\hat{\mathbf{k}}\cdot\hat{\mathbf{p}}}\sum_{m_{1},m_{2}}\sum_{m_{L},m_{S}}\sum_{m_{1}',m_{2}'}\sum_{m_{L}',m_{S}'}\\\nonumber &C_{1,\nicefrac{1}{2},S'}^{m_{1}',m_{2}',m_{S}'}C_{1,\nicefrac{1}{2},S}^{m_{1},m_{2},m_{S}}C_{L,S,J}^{m_{L},m_{S},M}C_{L',S',J}^{m_{L}',m_{S}',M}\\\nonumber
&{Y_{L'}^{m_{L}'}}^{*}(\hat{\mathbf{p}})Y_{L}^{m_{L}}(\hat{\mathbf{k}})\left(kY_{1}^{-m}(\hat{\mathbf{k}})+2pY_{1}^{-m}(\hat{\mathbf{p}})\right)(-1)^{m}\\\nonumber
&\sqrt{\frac{2}{3}}\sum_{m,m'}\sum_{k,q}\sqrt{2k+1}C_{1,1,1}^{m_{1},m,m'}C_{1,k,1}^{m_{1}',q,m'}C_{\nicefrac{1}{2},k,\nicefrac{1}{2}}^{m_{2},q,m_{2}'}\langle\nicefrac{1}{2}||T_{k}||\nicefrac{1}{2}\rangle
\end{align}



\noindent Integration over the angular variable can be carried out trivially by using (\ref{eq:AngularProjection2}) leaving a sum of products of Clebsch-Gordan coefficients . Then using symmetry properties of the Clebsch-Gordan coefficients we find.

\begin{align}
V_{JL}=&8\pi g^{{}^{3}\!S_{1}-{}^{3}\!P_{1}}\sqrt{3}C_{L',1,L}^{0,0,0}(-1)^{L-S-J}\frac{1}{kp}(kQ_{L'}(a)+2pQ_{L}(a))\times\\\nonumber
&\times\sum_{k}\sqrt{\bar{S}\bar{S'}\bar{L'}\bar{k}}\langle\nicefrac{1}{2}||T_{k}||\nicefrac{1}{2}\rangle
\left\{\begin{array}{ccc}
\nicefrac{1}{2} & k & \nicefrac{1}{2}\\
1 & S' & 1
\end{array}\right\}
\left\{\begin{array}{ccc}
1 & 1 & 1\\
\nicefrac{1}{2} & S' & S
\end{array}\right\}
\left\{\begin{array}{ccc}
L' & 1 & L \\
S & J & S'
\end{array}\right\}
\end{align}

\noindent in terms of 6-j symbols\cite{Strobel}.\newline\indent The sum over $k$ can be removed by use of the  identity\cite{Edmonds}. 

\begin{equation}
\label{eq:Edmonds}
\sum_{k}\bar{k}
\left\{\begin{array}{ccc}
\nicefrac{1}{2} & k & \nicefrac{1}{2}\\
1 & S' & 1
\end{array}\right\}
\left\{\begin{array}{ccc}
\nicefrac{1}{2}& \nicefrac{1}{2} & k\\
1 & 1 & j
\end{array}\right\}=\delta_{S'j}\frac{1}{\bar{S'}}
\end{equation}




\noindent yielding

\begin{align}
V_{JL}=&8\pi g^{{}^{3}\!S_{1}-{}^{3}\!P_{1}}\sqrt{6}C_{L',1,L}^{0,0,0}(-1)^{L-S-J}\sqrt{\bar{S}\bar{S'}\bar{L'}}\left(\frac{1}{2}\delta_{S'\nicefrac{1}{2}}+\delta_{S'\nicefrac{3}{2}}\right)\times\\\nonumber
&\times\left\{\begin{array}{ccc}
1 & 1 & 1\\
\nicefrac{1}{2} & S' & S
\end{array}\right\}
\left\{\begin{array}{ccc}
L' & 1 & L \\
S & J & S'
\end{array}\right\}\frac{1}{kp}(kQ_{L'}(a)+2pQ_{L}(a))
\end{align}

Using time reversal invariance and including our coefficients, as well as the isospin projection, we find the projection for the $\left[\boldsymbol{\mathcal{K}}(k,p)^{J}_{L'S',LS}\right]_{11}$ term is

\begin{align}
&-y_{d}g^{{}^{3}\!S_{1}-{}^{1}\!P_{1}}4\pi\sqrt{3}(-1)^{\nicefrac{3}{2}+2S+L-J}\delta_{S'\nicefrac{1}{2}}\sqrt{\bar{S}\bar{L}}C_{L,1,L'}^{0,0,0}
\left\{\begin{array}{ccc}
\nicefrac{1}{2} & 1 & S\\
L & J & L'
\end{array}\right\}
\frac{1}{kp}(kQ_{L'}(a)+2pQ_{L}(a))\\\nonumber
&-y_{d}g^{{}^{3}\!S_{1}-{}^{1}\!P_{1}}4\pi\sqrt{3}(-1)^{\nicefrac{3}{2}+2S'+L'-J}\delta_{S\nicefrac{1}{2}}\sqrt{\bar{S'}\bar{L'}}C_{L',1,L}^{0,0,0}
\left\{\begin{array}{ccc}
\nicefrac{1}{2} & 1 & S'\\
L' & J & L
\end{array}\right\}
\frac{1}{kp}(2kQ_{L'}(a)+pQ_{L}(a))\\\nonumber
&-y_{d}g^{{}^{3}\!S_{1}-{}^{3}\!P_{1}}8\pi\sqrt{6}C_{L',1,L}^{0,0,0}(-1)^{L-S-J}\sqrt{\bar{S}\bar{S'}\bar{L'}}\left(\frac{1}{2}\delta_{S'\nicefrac{1}{2}}+\delta_{S'\nicefrac{3}{2}}\right)
\left\{\begin{array}{ccc}
1 & 1 & 1\\
\nicefrac{1}{2} & S' & S
\end{array}\right\}
\left\{\begin{array}{ccc}
L' & 1 & L \\
S & J & S'
\end{array}\right\}\times\\\nonumber
&\times\frac{1}{kp}(kQ_{L'}(a)+2pQ_{L}(a))\\\nonumber
&-y_{d}g^{{}^{3}\!S_{1}-{}^{3}\!P_{1}}8\pi\sqrt{6}C_{L,1,L'}^{0,0,0}(-1)^{L'-S'-J}\sqrt{\bar{S}\bar{S'}\bar{L}}\left(\frac{1}{2}\delta_{S\nicefrac{1}{2}}+\delta_{S\nicefrac{3}{2}}\right)
\left\{\begin{array}{ccc}
1 & 1 & 1\\
\nicefrac{1}{2} & S & S'
\end{array}\right\}
\left\{\begin{array}{ccc}
L & 1 & L' \\
S' & J & S
\end{array}\right\}\times\\\nonumber
&\times\frac{1}{kp}(2kQ_{L'}(a)+pQ_{L}(a))
\end{align}


Now combining isospin with our spin projections we find the $\left[\boldsymbol{\mathcal{K}}(k,p)^{J}_{L'S',LS}\right]_{12}$ term projected out is

\begin{align}
&-8\pi\sqrt{3}(-1)^{1+S'+L-J}\delta_{S\nicefrac{1}{2}}y_{d}\left(\frac{2}{3}g^{{}^{1}\!S_{0}-{}^{3}\!P_{0}}_{(\Delta I=1)}-g^{{}^{1}\!S_{0}-{}^{3}\!P_{0}}_{(\Delta I=0)}\right)\times\\\nonumber
&\times C_{L',1,L}^{0,0,0}\sqrt{\bar{L'}\bar{S'}}\left(\frac{1}{2}\delta_{S'\nicefrac{1}{2}}+\delta_{S'\nicefrac{3}{2}}\right)
\left\{\begin{array}{ccc}
L' & 1 & L\\
S & J & S'
\end{array}\right\}\frac{1}{kp}(kQ_{L'}(a)+2pQ_{L}(a))\\\nonumber
&-y_{t}g^{{}^{3}\!S_{1}-{}^{1}\!P_{1}}4\pi\sqrt{3}(-1)^{\nicefrac{1}{2}+2S'+L-J}\delta_{S\nicefrac{1}{2}}\sqrt{\bar{L'}\bar{S'}}C_{L',1,L}^{0,0,0}
\left\{\begin{array}{ccc}
L' & 1 & L\\
S & J & S'
\end{array}\right\}
\frac{1}{kp}\left(2kQ_{L'}(a)+pQ_{L}(a)\right)\\\nonumber
&+y_{t}g^{{}^{3}\!S_{1}-{}^{3}\!P_{1}}24\pi\frac{1}{\sqrt{3}}(-1)^{S'-L-J}\delta_{S\nicefrac{1}{2}}\sqrt{\bar{S'}\bar{L'}}C_{L',1,L}^{0,0,0}
\left\{\begin{array}{ccc}
1 & 1 & 1\\
\nicefrac{1}{2} & S' & \nicefrac{1}{2}
\end{array}\right\}
\left\{\begin{array}{ccc}
L' & 1 & L\\
S & J & S'
\end{array}\right\}\frac{1}{kp}(2kQ_{L'}(a)+pQ_{L}(a))
\end{align}


Now using time reversal symmetry we see the projection of the $\left[\boldsymbol{\mathcal{K}}(k,p)^{J}_{L'S',LS}\right]_{21}$ term is

\begin{align}
&-8\pi\sqrt{3}(-1)^{1+S+L'-J}\delta_{S'\nicefrac{1}{2}}y_{d}\left(\frac{2}{3}g^{{}^{1}\!S_{0}-{}^{3}\!P_{0}}_{(\Delta I=1)}-g^{{}^{1}\!S_{0}-{}^{3}\!P_{0}}_{(\Delta I=0)}\right)\times\\\nonumber
&\times C_{L,1,L'}^{0,0,0}\sqrt{\bar{L}\bar{S}}\left(\frac{1}{2}\delta_{S\nicefrac{1}{2}}+\delta_{S\nicefrac{3}{2}}\right)
\left\{\begin{array}{ccc}
L & 1 & L'\\
S' & J & S
\end{array}\right\}\frac{1}{kp}(2kQ_{L'}(a)+pQ_{L}(a))\\\nonumber
&-y_{t}g^{{}^{3}\!S_{1}-{}^{1}\!P_{1}}4\pi\sqrt{3}(-1)^{\nicefrac{1}{2}+2S+L'-J}\delta_{S'\nicefrac{1}{2}}\sqrt{\bar{L}\bar{S}}C_{L,1,L'}^{0,0,0}
\left\{\begin{array}{ccc}
L & 1 & L'\\
S' & J & S
\end{array}\right\}
\frac{1}{kp}\left(kQ_{L'}(a)+2pQ_{L}(a)\right)\\\nonumber
&+y_{t}g^{{}^{3}\!S_{1}-{}^{3}\!P_{1}}24\pi\frac{1}{\sqrt{3}}(-1)^{S-L'-J}\delta_{S'\nicefrac{1}{2}}\sqrt{\bar{S}\bar{L}}C_{L,1,L'}^{0,0,0}
\left\{\begin{array}{ccc}
1 & 1 & 1\\
\nicefrac{1}{2} & S & \nicefrac{1}{2}
\end{array}\right\}
\left\{\begin{array}{ccc}
L & 1 & L'\\
S' & J & S
\end{array}\right\}\frac{1}{kp}(kQ_{L'}(a)+2pQ_{L}(a))
\end{align}


Now combining the spin with the isospin projections we find the projection of the $\left[\boldsymbol{\mathcal{K}}(k,p)^{J}_{L'S',LS}\right]_{22}$ term is

\begin{align}
-y_{t}g^{{}^{1}\!S_{0}-{}^{3}\!P_{0}}_{(\Delta I=0)}12\pi\sqrt{6}(-1)^{\nicefrac{1}{2}-L-J}\delta_{S'\nicefrac{1}{2}}\delta_{S\nicefrac{1}{2}}\sqrt{\bar{L'}}C_{L',1,L}^{0,0,0}
\left\{\begin{array}{ccc}
L' & 1 & L\\
S & J & S'
\end{array}\right\}\frac{1}{kp}(kQ_{L'}(a)+pQ_{L}(a))\\\nonumber
+y_{t}g^{{}^{1}\!S_{0}-{}^{3}\!P_{0}}_{(\Delta I=1)}8\pi\sqrt{6}(-1)^{\nicefrac{1}{2}-L-J}\delta_{S'\nicefrac{1}{2}}\delta_{S\nicefrac{1}{2}}\sqrt{\bar{L'}}C_{L',1,L}^{0,0,0}
\left\{\begin{array}{ccc}
L' & 1 & L\\
S & J & S'
\end{array}\right\}\frac{1}{kp}(kQ_{L'}(a)+pQ_{L}(a))\\\nonumber
\end{align}


\vspace{-.2cm}
\begin{acknowledgments}
I would like to thank Harald W. Griesshammer and L. Girlanda for useful discussions, Barry R. Holstein for guidance during this project, and U. van Kolck for useful input.  This work is supported in part by the National Science Foundation under Grant No. NSF-PHY 0855119
\end{acknowledgments}


\end{thebibliography}

\end{document}